\documentclass[twocolumn]{jpsj3}
\usepackage{bm}

\begin{document}

\def\bra{\langle}  \def\ket{\rangle}
\def\ketpsi{| \psi \rangle}


\title{
Necessity of Superposition of Macroscopically Distinct States\\ 
for Quantum Computational Speedup
}
\author{
Akira \textsc{Shimizu}\thanks{E-mail address: shmz@ASone.c.u-tokyo.ac.jp},
Yuichiro \textsc{Matsuzaki}$^{1}$\thanks{E-mail address: matsuzaki.yuichiro@lab.ntt.co.jp},
and
Akihisa \textsc{Ukena}
}
\inst{
Department of Basic Science, University of Tokyo, 
3-8-1 Komaba, Tokyo 153-8902, Japan\\
$^{1}$
NTT Basic Research Laboratories, NTT Corporation, Kanagawa 243-0198, Japan
}
\date{\today}

\def\runauthor{\sc A. Shimizu, Y. Matsuzaki and A. Ukena}

\abst{
For quantum computation, we investigate the conjecture
that the superposition of macroscopically distinct states is necessary 
for a large quantum speedup.
Although this conjecture was supported for 
a circuit-based quantum computer performing Shor's factoring algorithm
[A. Ukena and A. Shimizu, Phys.\ Rev. A {\bf 69} (2004) 022301), 
it needs to be generalized for it to be applicable to 
a large class of algorithms
and/or other models such as measurement-based quantum 
computers.
To treat such general cases, 
we first generalize the indices for 
the superposition of macroscopically distinct states.
We then generalize the conjecture, 
using the generalized indices,
in such a way that it is unambiguously applicable 
to general models if a
quantum algorithm achieves exponential speedup.
On the basis of this generalized conjecture, 
we further extend the conjecture to Grover's quantum search algorithm,
whose speedup is large but quadratic. 
It is shown that this extended conjecture is also correct.
Since Grover's algorithm is a representative algorithm 
for unstructured problems, the present result further supports the conjecture.
}

\kword{
quantum computer, entanglement, macroscopic quantum system
}

\maketitle

\section{Introduction}\label{sec:intro}

We consider quantum speedup for solving 
computational problems of size $L$ bits, such as
the factoring problem (for which $L$ is the size of the number to be factored)
and the search problem ($L$ is the size of the solution space).
In the well-known quantum algorithms of 
Shor \cite{Shor,EJ} and Grover \cite{Grover}, 
such problems are solved using quantum computers
whose number of qubits ${\cal L} \geq L$ 
\cite{Shor,EJ,Grover,NC}.
Since quantum speedup becomes relevant for large $L$, 
such quantum computers are many-body quantum systems
with a large number of qubits ${\cal L}$.
Since there are many types 
(and corresponding measures or indices)
of entanglement for many-body systems
\cite{Bennett,miyake,SM02,Meyer_Wallach,Stockton,Syljuasen,Verstraete,
MSS05,SS05}, 
it is interesting to determine which types 
of entanglement
are relevant to a large quantum speedup
over classical computations 
\cite{PP,JL,Vidal,US04,OL,SSB04,SSB05,conjecture1,conjecture2}.

This issue has been studied extensively, particularly in 
Shor's factoring algorithm \cite{Shor,EJ}
and Grover's quantum search algorithm \cite{Grover}.
For example, Parker and Plenio demonstrated that bipartite 
entanglement as measured by the logarithmic negativity 
is an intrinsic part of Shor's algorithm \cite{PP}.
Shimoni et al. showed that 
highly entangled states are generated in both algorithms \cite{SSB04,SSB05}.
Or\'us and Latorre 
studied the scaling of entanglement in three algorithms including 
Shor's and Grover's \cite{OL}.

For general algorithms, 
a few necessary conditions were derived for computational speedup
over classical computations.
Jozsa and Linden showed that, 
for exponential speedup,
a state that 
cannot be factored into a direct 
product of states of at most a constant number of qubits
is necessary \cite{JL}.
Vidal showed that a necessary condition for exponential 
speedup is that 
the amount of bipartite entanglement 
between one part and the rest of the qubits
should increase with $L$ \cite{Vidal}.

We note that one can get a stricter condition by taking the product of 
these and other necessary conditions, 
which may be obtained by studying other types 
of entanglement.
Such a stricter condition would lead to 
a deeper understanding of quantum computations.
Hence, it is important to seek more conditions that are 
necessary for quantum computational speedup.

As a possible necessary condition, 
one of the authors conjectured that 
the superposition of macroscopically distinct states is necessary for 
quantum computational speedup (refs.~\citen{conjecture1, conjecture2} and \S\ref{sec:conj}).
Although the `superposition of macroscopically distinct states'
was only ambiguously defined until 
recently, a clear definition and 
the corresponding index $p$ ($1 \leq p \leq 2$) for pure states
were proposed in refs.~\citen{SM02,MSS05}, according to which 
a pure state has a superposition of macroscopically distinct states
if $p=2$. 
The generalization to mixed states was made in ref.~\citen{SM05},
in which $p$ is generalized to an index $q$ ($1 \leq q \leq 2$);
a mixed state has a superposition of macroscopically distinct states
if $q=2$. 
For pure states, $p=2$ implies $q=2$ and vice versa \cite{SM05},
whereas $p$ is undefined for mixed states.
 
A superposition of macroscopically distinct states 
is an entangled state.
However, 
its entanglement cannot be quantified well by 
bipartite entanglement, which was studied in 
previous works \cite{JL,Vidal}.
[Hence, it was called `macroscopic entanglement' in 
refs.~\citen{SM05,SS05,MSS05} and \citen{MS06}. 
However, we do not use this term in this paper
because the same term is used in several other senses by other authors.]
For example, 
some pure states with $p=2$ (such as the GHZ state) 
have a very small bipartite entanglement, 
whereas some other states with $p=1$ (such as energy eigenstates of
many-body chaotic systems)
have very large bipartite entanglement \cite{MSS05,SS05}. 
Therefore, the 
{\em simultaneous} requirement (for quantum computational speedup)
of the superposition of macroscopically distinct states
{\em and} large bipartite entanglement
is much stronger than the requirement of either one of them.

For a circuit-based quantum computer 
performing Shor's factoring algorithm \cite{Shor,EJ},
we obtained results that support
the above conjecture in a previous paper \cite{US04}.
It is interesting to study the correctness of the conjecture 
in other algorithms and/or other models
(such as measurement-based quantum computers).
To explore such general cases unambiguously, however, 
the conjecture needs to be generalized.
For example, 
quantum states in quantum computers are not only {\em inhomogeneous} 
but also {\em dependent on instances} (i.e., 
different for different questions of a given problem).
Since the indices $p$ and $q$ 
assume a family of similar states that are spatially homogeneous,
the conjecture (which is based on $p$ or $q$) is not 
strictly applicable 
to such a general family of states,
in its original form.

Furthermore, 
since Shor's algorithm is a representative quantum algorithm
for solving structured problems \cite{NC}, 
it is very interesting to study whether
the conjecture is correct in the case of 
quantum algorithms for solving {\em unstructured} problems.
However, 
the quantum speedup achieved by 
Grover's quantum search algorithm \cite{Grover}, 
which is a representative algorithm 
for unstructured problems \cite{NC},
is not exponential but quadratic.
The conjecture, in its original form, does not assume 
speedup of such a degree.

In this paper, 
we first generalize the indices $p$ and $q$
to treat general algorithms and models.
We then generalize the conjecture, 
using the generalized indices,
in such a way that it is unambiguously applicable 
to general models if a
quantum algorithm achieves exponential speedup.
On the basis of this generalized conjecture, 
we further extend the conjecture to the quadratic speedup 
of Grover's quantum search algorithm.
It is shown that this extended conjecture is also correct
for Grover's algorithm.
To show details of the evolution of the 
superposition of macroscopically distinct states, 
we also perform numerical simulations of a quantum computer that performs 
Grover's quantum search algorithm.

This paper is organized as follows.
In \S\ref{sec:p.and.q}, we generalize the indices $p$ and $q$.
Section \ref{sec:conj} is devoted to 
generalizing and further extending the conjecture.
Analytic results for Grover's quantum search algorithm
are given in \S\ref{sec:Grover-analytic}, where 
we will prove that the extended conjecture is correct
for Grover's algorithm.
We present results of 
numerical simulations of a quantum computer that performs 
Grover's quantum search algorithm
in \S\ref{sec:Grover-numerical}.
Discussions and summary are given in \S\ref{sec:dc}.

\section{Indices of Superposition of Macroscopically Distinct States}
\label{sec:p.and.q}

The indices of 
the superposition of macroscopically distinct states 
were proposed and studied 
for pure states in refs.~\citen{SM02,MSS05,MS06},
and for mixed states in ref.~\citen{SM05}.
To study these indices for states in quantum computers,
we here generalize their definitions because, 
as will be illustrated explicitly in \S\ref{sec:Grover-analytic} and 
\S\ref{sec:Grover-numerical}, 
quantum states in quantum computers are not only inhomogeneous 
but also dependent on instances.
Here, an {\em instance} is a particular question of a given problem.
The physical meanings and implications 
of the indices will also be described briefly 
in this section.

\subsection{Index for a family of pure states}
\label{sec:p}

For the index $p$ of 
the superposition of macroscopically distinct states,
which was proposed and studied in refs.~\citen{SM02,MSS05} and \citen{MS06},
the main point is described in Appendix \ref{app:p}.
We here generalize it.

Consider a quantum system of size $L$.
Let $| \psi_\nu(L) \rangle$'s be its pure states, 
which are labeled by the index $\nu$, for example, 
as $| \psi_1(L) \rangle, | \psi_2(L) \rangle, | \psi_3(L) \rangle, \cdots$.
For each $L$, the range of $\nu$ is given, say, 
as $\nu=1, 2, \ldots, 2^L$.
In a quantum computer that solves a decision problem, 
$L$ corresponds to the size of a certain register, 
which is usually proportional to the size of the input
of the problem, and $\nu$ labels various inputs
(see later sections). 
We do {\em not} assume that 
$\langle \psi_\nu(L) | \psi_{\nu'}(L) \rangle = 0$
for $\nu \neq \nu'$.
We consider a family $F$ of $| \psi_\nu(L) \rangle$'s, i.e., 
\begin{equation}
F \equiv 
\{ | \psi_\nu(L) \rangle \ | \mbox{ all $\nu$'s, all $L$} \},
\label{def:F}\end{equation}
which we abbreviate to $\{ | \psi_\nu(L) \rangle \}_{L,\nu}$.

In general, a quantum computer consists of a large number of small 
quantum systems, such as qubits, which are distributed spatially.
We call each small quantum system a {\em site},
and an operator acting on a single site a {\em local operator}.
To avoid mathematical complexities, 
we limit ourselves to the case where 
every local operator 
is bounded (i.e., its norm is finite).

Let $\hat a(l)$ be a local operator on site $l$.
We normalize it as $\| \hat a(l) \|=1$.
Although this normalization condition 
might look too restrictive at first sight, 
it actually imposes only a weak restriction 
that makes the 
maximization 
operation in eq.~(\ref{def:p}) well-defined \cite{note:a(l)},
as discussed in Appendix \ref{app:why|a|=1}.

Note that we can use either $\| \hat a \|_E$ or $\| \hat a \|_H$
as the operator norm $\| \hat a \|$,
where
\begin{eqnarray}
\| \hat a \|_E &\equiv& 
\sup_{\| | \psi \rangle \|=1} \| \hat a | \psi \rangle \|,
\label{norm.E}
\\
\| \hat a \|_H &\equiv& \sqrt{(\hat a, \hat a)},
\label{norm.H}\end{eqnarray}
and
$(\hat a, \hat b) \equiv {\rm Tr} (\hat a^\dagger \hat b)$
denotes the inner product of operators $\hat a$ and $\hat b$.
In fact, 
both definitions give the same index $p$,
because 
$\| \hat a(l) \|_E \leq \| \hat a(l) \|_H \leq \sqrt{d} \| \hat a(l) \|_E$
and
$\| \hat a(l) \|_H/\sqrt{d} \leq \| \hat a(l) \|_E \leq \| \hat a(l) \|_H$,
where $d$ is the dimension of the Hilbert space of a single site.

By the same symbol $\hat a(l)$,
we also denote 
$\displaystyle \hat a(l) \otimes \bigotimes_{l' (\neq l)}\hat 1(l')$,
which is an operator on the Hilbert space ${\cal H}$ of the total system,
where $\hat 1(l')$ is the identity operator acting on site $l'$.
Using this notation, 
we define an {\em additive operator} $\hat{A}(L)$
as the sum of local operators \cite{SM02,SM05}:
\begin{equation}
\hat{A}(L) = \sum_{l=1}^L \hat a(l) \quad (\| \hat a(l) \|=1). 
\label{def:A}\end{equation}
Here, we do {\em not} assume that $\hat a(l')$ ($l' \neq l$) 
is a spatial translation of $\hat a(l)$.

To simplify the notation, 
we express the expectation value in 
$|\psi_\nu(L) \rangle$
as $\langle \cdot \rangle_{L \nu}$.
We also use the symbols $O, \Omega$ and $\Theta$ to 
describe asymptotic behaviors according to
ref.~\citen{NC}, as summarized in Appendix \ref{app:notation}.
Furthermore, as described in Appendix \ref{app:notation}, 
a family $\{ f_\nu(L) \}_{L,\nu}$ of non-negative functions of $L$
is said to be $\overline{\Theta}(g(L))$ 
if $f_\nu(L)$ is $\Theta(g(L))$ 
{\em for almost every $\nu$},
i.e., apart from possible exceptional
$\nu$'s whose measure (i.e., the number of such $\nu$'s
divided by the total number of $\nu$'s) vanishes as $L$
goes to infinity.

For each state $| \psi_\nu(L) \rangle$, 
consider the fluctuation 
$\langle \Delta \hat A(L)^\dagger \Delta \hat A(L) \rangle_{L \nu}$,
where 
$\Delta \hat A(L) \equiv \hat A(L) - \langle \hat A(L) \rangle_{L \nu}$.
Its magnitude 
depends on $\hat{A}(L)$, i.e., on the choice of $\hat a(l)$'s.
Since $\| \hat A(L) \|$ is upper-bounded, there exists a maximum value, 
$\max_{\hat A(L)} 
\langle \Delta \hat A(L)^\dagger \Delta \hat A(L) \rangle_{L \nu}
$.
The maximum value is taken for some additive operator
$\hat A_{\max}(L)$, 
which we call the {\em most fluctuating additive operator}.
Using the maximum value, which depends on $L$ and $\nu$,
we define the index $p$ of the family 
$F$ ($= \{ | \psi_\nu(L) \rangle \}_{L,\nu}$) 
as the positive number (if it exists) that satisfies
\begin{equation}
\max_{\hat A(L)} \ 
\langle \psi_\nu(L) | \Delta \hat A(L)^\dagger \Delta \hat A(L) 
|\psi_\nu(L) \rangle
= \overline{\Theta}(L^{p}).
\label{def:p}
\end{equation}
Note that $p$ does not necessarily exist for a general family.
If $p$ exists for a given family, 
we can show (see Appendix \ref{appendix_range}) that 
\begin{equation}
1 \leq p \leq 2.
\end{equation}
When a family $\{ | \psi_\nu(L) \rangle \}_{L,\nu}$
has $p=2$ (or $p=1$, etc), 
we also simply say that `almost every state 
$| \psi_\nu(L) \rangle$ has $p=2$ (or $p=1$, etc).'

The present definition of $p$
contains those of the previous works,
refs.~\citen{SM02,MSS05,SS05} and \citen{MS06},
as special cases.
In fact, 
refs.~\citen{SM02,MSS05} and \citen{MS06} treated homogeneous states, 
for which $| \psi(L+1) \rangle$ 
was simply defined as an enlarged state of $| \psi(L) \rangle$.
For example, when 
$| \psi(L) \rangle = |00 \cdots 0 \rangle + |11 \cdots 1 \rangle$ 
with $L$ sites,  
then 
$| \psi(L+1) \rangle = |000 \cdots 0 \rangle + |111 \cdots 1 \rangle$ 
with $(L+1)$ sites.
According to the present general definition of $p$, 
such a case corresponds to a family 
whose members for each $L$ are identical, i.e., 
$| \psi_\nu(L) \rangle = | \psi(L) \rangle$ for all $\nu$.
Moreover, ref.~\citen{SS05} treated 
energy eigenstates of homogeneous chaotic systems.
Since {\em each} energy eigenstate is {\em inhomogeneous} spatially, 
this case is different from the case of homogeneous states.
According to the present definition of $p$, 
it corresponds to a family composed of 
all energy eigenstates in a certain energy interval.
Therefore, 
$p$ defined here is a natural generalization of 
$p$ of refs.~\citen{SM02,MSS05,SS05} and \citen{MS06}.

The physical meaning of the present $p$ is basically the same as 
the $p$ of the previous works mentioned. 
That is, 
as explained in Appendix \ref{app:p}, 
almost every state of a family with $p=2$ 
is
a superposition of states with macroscopically distinct values of
some additive operator(s) \cite{SM02,MSS05,MS06,SS05}.
We call such an additive operator(s) a 
{\em macroscopically-fluctuating additive operator(s)} of 
the state (or family).
For example, when 
$| \psi_\nu(L) \rangle = |00 \cdots 0 \rangle + |11 \cdots 1 \rangle$ 
for all $\nu$ then its macroscopically-fluctuating additive operator
is $\sum_l \hat{\sigma}_z(l)$,
which corresponds to the $z$ component of 
the total magnetization of a magnetic substance.
Since additive operators are macroscopic dynamical variables \cite{SM02,SM05},
two (or more) states are macroscopically distinct from each other
if they have macroscopically distinct values of an additive operator.
Therefore, one can definitely state that 
a state with $p=2$ is 
a superposition of macroscopically distinct states.

In the present general definition of $p$, 
a macroscopically-fluctuating additive operator(s)
of a given family
can be different for different $L$ and $\nu$,
unlike in the case of 
homogeneous states treated in refs.~\citen{SM02,MSS05} and \citen{MS06}.
Therefore, 
we can say the following:
{\em 
For a given family 
of pure states,
if there exists a family 
of additive operators
\begin{equation}
\{ \hat A_\nu(L) \ | \mbox{ all $\nu$'s, all $L$} \}
\end{equation}
such that 
\begin{equation}
\langle \Delta \hat A_\nu(L)^\dagger \Delta \hat A_\nu(L) \rangle_{L \nu}
= \overline{\Theta}(L^{2}),
\label{p=2_some_A}
\end{equation}
then the family has $p=2$. 
}
This means that 
almost all states of the family are
superpositions of states that have macroscopically distinct values
of these additive operator(s). 

Note that there often exist two or more families of such
operators, even for a family of homogeneous states \cite{MSS05,MS06}.

\subsection{Index for a family of mixed states}
\label{sec:q}

The index $p$, which is defined only for pure states, is sufficient
for the concrete analyses in \S\ref{sec:Grover-analytic} and 
\S\ref{sec:Grover-numerical}.
However, to state our conjecture in a general form, 
we need a more general index which is applicable to mixed states.
For example, in the measurement-based quantum computation \cite{mbqc1,mbqc2},
although the entire qubits are entangled as a cluster state, only a 
small number of qubits (called `logical qubits') has information on the computation, whereas the other qubits are prepared as ancillary qubits. 
In this case, to determine whether a superposition of macroscopically distinct states appears during the computation, we need to calculate an index for the reduced density matrix of the logical qubits \cite{one-way.Grover}.
Fortunately, the generalization of $p$ 
to mixed states has been carried out in ref.~\citen{SM05},
in which the generalized index $q$ 
was proposed
for homogeneous mixed states.
Here, we generalize it to families of more general 
mixed states in order to apply $q$ to a large class of quantum computers.

For an additive operator $\hat A(L)$, 
as given by eq.~(\ref{def:A}),
and 
a projection operator $\hat \eta$ on ${\cal H}$, 
satisfying $\hat \eta^2 = \hat \eta$,
we define the Hermitian operator
\begin{equation}
\hat C_{\hat A \hat \eta} \equiv 
[ \hat A, [ \hat A, \hat \eta ] ]
=
\hat A^2 \hat \eta - 2 \hat A \hat \eta \hat A + \hat \eta \hat A^2.
\label{eq:C}\end{equation}
For a family $\{ \hat \rho_\nu(L) \}_{L,\nu}$ of mixed states,
which are not necessarily homogeneous spatially, 
we define the index $q$ 
as the positive number (if it exists) that satisfies
\begin{equation}
\max \left\{ 
L,\
\max_{\hat A, \hat \eta} \langle C \rangle_{L \nu}
\right\} = \overline{\Theta}(L^q),
\label{def:q}\end{equation}
where 
$\max_{\hat A, \hat \eta}$ is taken over all 
possible choices of $\hat A$ and $\hat \eta$.
If this $q$ exists for a given family,
we can show (by slightly generalizing the proof in ref.~\citen{SM05})
that 
\begin{equation}
1 \leq q \leq 2.
\end{equation}
We say that the family $\{ \hat \rho_\nu(L) \}_{L,\nu}$ of mixed states
is a superposition of macroscopically distinct states 
if 
$q$ exists and $q=2$.
We also say that almost every $\hat \rho_\nu(L)$
of such a family 
is a superposition of macroscopically distinct states.
For pure states, 
this is consistent with the corresponding statement based on $p$,
because we can show (following the proof in ref.~\citen{SM05})
that, for pure states, $p=2$ implies $q=2$ and vice versa.

The case of ref.~\citen{SM05} corresponds to the special case where 
$\hat \rho_\nu(L)$'s are homogeneous and independent of $\nu$,
and 
$\hat \rho_\nu(L+1)$ is simply an enlarged state of $\hat \rho_\nu(L)$. 
In such a case, 
$q$ defined here reduces to $q$ of
ref.~\citen{SM05}.

\subsection{Properties of states with $p=2$ or $q=2$}
\label{sec:Pofpq}

As application of the general theory in ref.~\citen{SM02}, 
the index $p$ was studied 
for many-magnon states in ref.~\citen{MSS05},
for energy eigenstates of many-body chaotic systems in ref.~\citen{SS05},
and for typical many-body states in ref.~\citen{MS06}.
A comparison of $p$ with a measure of bipartite entanglement was also made 
in these references.
Most importantly, 
many states
 were found (such as energy eigenstates of a chaotic system \cite{SS05})
 such that they are almost maximally 
entangled in the bipartite measure but their $p$ is minimum, $p=1$.
Many other states (such as the GHZ state) are also found 
such that their $p$ is maximum, $p=2$, but
their bipartite entanglement (as measured by the von Neumann entropy of the 
reduced density operator) is small.
Therefore, the aspect of entanglement detected by $p$ or $q$
is completely different from that detected by the bipartite measure.
Furthermore, 
it has been shown in ref.~\citen{SM05}
that a family of 
states with $q=2$ has a strong $L$-point correlation,
which is $\Theta(L)$ times larger than that of 
any separable state.
Note that any measure of bipartite entanglement cannot detect
such a strong $L$-point correlation for mixed states.
On the other hand, 
the index $q$ does not detect 
entanglement generated by a small number of Bell pairs, 
whereas measures of bipartite entanglement do.
These findings demonstrate that 
{\em the index $q$ is complementary to
the measures of bipartite entanglement}.

Note that $p$ is calculated from {\em two-point} correlations
because the fluctuation of an additive operator is the sum of
two-point correlations:
\begin{equation}
\langle \Delta \hat A(L)^\dagger \Delta \hat A(L) \rangle_{L \nu}
=
\sum_{l=1}^L \sum_{l'=1}^L
\langle \Delta \hat a(l)^\dagger 
\Delta \hat a(l') \rangle_{L \nu}.
\end{equation}
However, this does {\em not} mean that 
$p$ is related {\em only} to two-point correlations,
because, as mentioned above, pure states with $p=2$ have $q=2$,
which means a strong $L$-point correlation.
That is, 
given the knowledge that a family consists of pure states,
one can say that if the family has $p=2$ then it has a strong $L$-point correlation.

It was shown in ref.~\citen{SM02} that 
$p$ is directly related to fundamental stabilities
of many-body states
against decoherence and local measurements.
Regarding decoherence by weak noises, 
it was shown that, for any state 
with $p=1$, its decoherence rate $\Gamma$ by {\em any} noise
{\em never} exceeds $O(L)$
[if the interaction between the noise and the system 
satisfies the locality condition, i.e., if it is the sum of 
local interactions].
For a state with $p=2$, on the other hand, 
it is possible in principle 
to construct a noise or environment 
that makes $\Gamma$ of the state $\Theta(L^2)$.
However, this does {\em not} necessarily mean that 
such a fatal noise or environment does exist in real physical systems; 
it depends on the physical situation \cite{SM02}.
A more fundamental stability is the 
stability against local measurements, which was proposed and 
defined in refs.~\citen{SM02} and \citen{Mtzk.cp}.
From the theorems proved in these references 
we can say that a state with $p=2$ is {\em unstable}
against local measurements, i.e., 
there exists a {\em local} observable such that measuring it
changes the state drastically \cite{2p-corr}.

Furthermore, quite a singular property was proved rigorously
in ref.~\citen{SM02}; 
any {\em pure} state with $p=2$ in a {\em finite} system of size $L$ 
does {\em not} approach a pure state in an {\em infinite} 
system as $L \to \infty$. 
For readers who are not familiar with the quantum theory of
infinite systems \cite{Haag}, 
we give a brief explanation of this fact in Appendix \ref{app:pure2mixed}.

These observations indicate that states with $p=2$ or $q=2$ 
are extremely anomalous many-body states.
This led to the conjecture in refs.~\citen{conjecture1} 
and \citen{conjecture2},
which will be generalized 
in \S\ref{sec:conj} of the present paper.

\subsection{Efficient method of identifying 
superposition of macroscopically distinct states}
\label{sec:VCM}

The evaluation of a measure or index of entanglement often becomes intractable
for large $L$.
Fortunately, this is not the case for $p$ because
there is an efficient method of calculating $p$ 
\cite{MSS05,SS05,MS06}.
Since this method assumed homogeneous states, 
we here generalize it to study general families of pure states.

In this subsection,
the dimension $d$ of the local Hilbert space is arbitrary, 
and we employ $\| \hat a \|_H$ defined by eq.~(\ref{norm.H})
as the operator norm $\| \hat a \|$.

Let $\{ \hat b_0(l), \hat b_1(l), \cdots, \hat b_D(l) \}$ be a complete 
orthonormal basis set of operators on site $l$, 
where $\hat b_0(l) = \hat 1(l)/\sqrt{d}$ and
$D \equiv d^2-1$.
We expand $\hat a(l)$ as 
\begin{equation}
\hat a(l) = \sum_{\alpha=0}^D c_{l \alpha} \hat b_\alpha(l),
\quad \sum_{\alpha=0}^D |c_{l \alpha}|^2 = 1,
\label{eq:a=cb}\end{equation}
where the latter equality comes from $\| \hat a(l) \|_H = 1$.
Let
$\Delta \hat a(l) 
\equiv \hat a(l) - \langle \hat a(l) \rangle_{L \nu}$
and
$\Delta \hat b_\alpha(l) 
\equiv \hat b_\alpha(L) - \langle \hat b_\alpha(L) \rangle_{L \nu}$.
Note that 
$\Delta \hat b_\alpha(l)$'s, unlike $\hat b_\alpha(l)$'s,
 are {\em not} necessarily orthogonal to each other.

Since $\Delta \hat b_0(l) = 0$, 
terms with $\alpha=0$ do not contribute to 
$\displaystyle \Delta \hat A(L) = \sum_l \Delta \hat a(l)$, i.e., 
it can be expanded as
\begin{equation}
\Delta \hat A(L)
=\sum_{l=1}^L \sum_{\alpha =1}^D c_{l \alpha} \Delta \hat b_\alpha(l),
\quad \sum_{\alpha=1}^D |c_{l \alpha}|^2 \leq 1.
\label{DA:sum_leq_1}
\end{equation}
As a result, 
$\sum_l \sum_{\alpha=1}^D |c_{l \alpha}|^2 \leq L$.
Since this normalization is not convenient, 
we temporarily consider another operator $\Delta \hat A'$
whose expansion coefficients are {\em normalized}; 
\begin{equation}
\Delta \hat A'(L)
=
\sum_{l=1}^L \sum_{\alpha =1}^D c'_{l \alpha} \Delta \hat b_\alpha(l),
\quad \sum_{l=1}^L \sum_{\alpha=1}^D |c'_{l \alpha}|^2 =L.
\label{DA:sum=L}\end{equation}
Here, we do {\em not} require that 
$\sum_{\alpha=1}^D |c'_{l \alpha}|^2 = 1$ for every $l$.
The fluctuation of such an operator is calculated as
\begin{equation}
\langle \Delta \hat A'(L)^\dagger \Delta \hat A'(L) \rangle_{L \nu}
=
\sum_{l,l'=1}^L \sum_{\alpha,\alpha' =1}^D
c^{\prime *}_{l \alpha} 
c'_{l' \alpha'} 
V_{l \alpha, l' \alpha'}(L,\nu)
\end{equation}
for each $| \psi_\nu(L) \rangle$. Here,
for $\alpha, \alpha'=1, 2, \cdots, D$ and 
$l, l'= 1, 2, \cdots, L$, we have defined
\begin{equation}
V_{l \alpha, l' \alpha'}(L,\nu)
\equiv
\langle \Delta \hat b^\dagger_\alpha(l) 
\Delta \hat b_{\alpha'}(l') \rangle_{L \nu},
\label{def:VCM}\end{equation}
which can be regarded as elements of 
a $DL \times DL$ Hermitian matrix, 
which we call the {\em variance-covariance matrix} (VCM).
Therefore, for each $| \psi_\nu(L) \rangle$, 
\begin{equation}
\max_{\hat A'(L)} 
\langle \Delta \hat A'(L)^\dagger \Delta \hat A'(L) \rangle_{L \nu}
=e_{\rm max}(L,\nu) L,
\label{eq:maxA'}\end{equation}
where $e_{\rm max}(L,\nu)$ is the maximum eigenvalue of the VCM.
This and eq.~(\ref{def:p}) suggest that the following index $p_e$ should be 
useful:
\begin{equation}
e_{\rm max}(L,\nu)
= \overline{\Theta}(L^{p_e-1}).
\label{def:p_e}
\end{equation}
In fact, we can show (see Appendix \ref{app:C=O(L^0)}) that 
{\em if $p_e=2$ then $p=2$ and vice versa}.
For a family of homogeneous states, in particular, 
we can show that $p=p_e$ for every value of $p$ 
(see the last paragraph of Appendix \ref{app:C=O(L^0)}).
Hence, {\em one can identify states with $p=2$ 
by calculating $e_{\rm max}(L,\nu)$}.
This can be carried out within the time Poly$(L)$ because the VCM is a
$DL \times DL$ 
matrix.

Furthermore, 
if $p_e=2$, 
we can find a macroscopically-fluctuating additive operator
from the eigenvector(s) 
$\{ c^{\prime \, \nu}_{l \alpha \, {\rm max}}(L) \}$ 
of the VCM corresponding to $e_{\rm max}(L,\nu)$,
as follows.
If we normalize $\{ c^{\prime \, \nu}_{l \alpha \, {\rm max}}(L) \}$ as 
\begin{equation}
\sum_{l=1}^L \sum_{\alpha=1}^D 
|c^{\prime \, \nu}_{l \alpha \, {\rm max}}(L)|^2 =L,
\label{normalization:c'max}\end{equation}
then the operator 
\begin{equation}
\Delta \hat{A}^{\prime \nu}_{\rm max}(L)
\equiv
\sum_{l=1}^L \sum_{\alpha =1}^D 
c^{\prime \, \nu}_{l \alpha \, {\rm max}}(L)
\Delta \hat b_\alpha(l)
\label{DAmax}\end{equation}
takes the form 
of eq.~(\ref{DA:sum=L}), and it fluctuates macroscopically:
\begin{equation}
\langle \Delta \hat{A}^{\prime \nu \dagger}_{\rm max}(L) 
\Delta \hat{A}^{\prime \nu}_{\rm max}(L) \rangle_{L \nu}
= e_{\rm max}(L,\nu) L
= \overline{\Theta}(L^{2}).
\label{fluc:DAmax}\end{equation}
As shown in Appendix \ref{app:C=O(L^0)}, if we put
\begin{equation}
C_\nu (L) \equiv 
\max_l \left( \sum_{\alpha=1}^D 
\left| c^{\prime \, \nu}_{l \alpha \, {\rm max}}(L) \right|^2 \right),
\label{Cnu(L)}\end{equation}
then $C_\nu (L) = \overline{\Omega}(L^0)$ if $p_e=2$.
Therefore, the operator
[which clearly takes the form of eq.~(\ref{DA:sum_leq_1})]
\begin{equation}
\Delta \hat{A}^{\nu}_{\rm max}(L)
\equiv
\sum_{l=1}^L \sum_{\alpha =1}^D 
c^\nu_{l \alpha \, {\rm max}}(L)
\Delta \hat b_\alpha(l),
\label{DAmax:normalized}\end{equation}
where
$c^\nu_{l \alpha \, {\rm max}}(L)
\equiv
c^{\prime \, \nu}_{l \alpha \, {\rm max}}(L)
/\sqrt{C_\nu (L)}$,
also fluctuates macroscopically:
\begin{eqnarray}
\langle \Delta \hat{A}^{\nu \dagger}_{\rm max}(L) 
\Delta \hat{A}^{\nu}_{\rm max}(L) \rangle_{L \nu}
&=& e_{\rm max}(L,\nu) L/C_\nu (L)
\nonumber\\
&=& \overline{\Theta}(L^{2}).
\label{fluc:DAmax:normalized}
\end{eqnarray}
We can then construct an additive operator 
$\hat{A}^\nu_{\rm max}(L)$
easily from 
$\Delta \hat{A}^\nu_{\rm max}(L)$,
by going from eq.~(\ref{DA:sum_leq_1}) back to eq.~(\ref{eq:a=cb}).
Although $\hat{A}^\nu_{\rm max}(L)$ is not uniquely determined 
from $\Delta \hat{A}^\nu_{\rm max}(L)$
(as discussed in Appendix \ref{app:why|a|=1}),
this nonuniqueness does not cause any difficulty 
because $p$ is defined only through the fluctuation.

\section{Conjecture on quantum computation}\label{sec:conj}

The conjecture in refs.~\citen{conjecture1} and \citen{conjecture2}
is roughly that 
the superposition of macroscopically distinct states  
is {\em necessary} for a large quantum speedup.
We now generalize it 
to treat a large class of algorithms and models.

We consider decision problems
because 
most computational problems can be reduced, with polynomial overheads, 
to some decision problems.
The number of bits or qubits in a computer is allowed to be Poly$(L)$,
where $L$ denotes the size of the input measured in bits.
To be definite, 
we assume that 
a quantum computer is composed of qubits (i.e., two-level quantum systems), 
which are separated spatially from each other.
That is, 
if we use the terms of the general discussion in the previous section, 
each qubit is located on its own site.

We consider the time complexity of problems,
allowing both quantum and classical algorithms
to have a bounded probability of error.
In doing so, we assume a classical 
probabilistic Turing machine as a 
counterpart of a quantum computer.

\subsection{Exponential speedup}
\label{sec:def-speedup}

To establish notation and to exclude possible ambiguity,  
we first define exponential speedup, according to convention, as follows.

We consider problems that are {\em not} in 
BPP (Bounded-error Probabilistic Polynomial time).
For a given (decision) problem, 
we denote an instance (input) by $i(L,\nu)$, 
where $\nu$ labels different instances (inputs) of size $L$.
The computational time depends not only on $L$ but also on $\nu$.
For a given classical computer $C$ and a given quantum computer $Q$,
let $T_C(L,\nu)$ and $T_Q(L,\nu)$, respectively, be the computational time 
for an instance $i(L,\nu)$,
with a bounded probability of error being allowed.

Since we are considering a decision problem that is 
{\em not} in BPP, 
for any classical computer $C$ there exists a set of infinitely 
many instances that cannot be solved in polynomial time.
That is, 
\begin{eqnarray}
&& T_C(L,\nu) > \mbox{Poly}(L)
\ \mbox{for some infinitely} 
\nonumber\\
&& \qquad \mbox{many instances in } \{ i(L,\nu) \}_{L,\nu}.
\label{eq:TC>P}
\end{eqnarray}
Here, $T_C(L,\nu) > \mbox{Poly}(L)$ means that $T_C(L,\nu) = \Omega(P(L))$
for any polynomial $P(L)$.
For a quantum computer $Q$ solving such a problem, 
we say $Q$ achieves 
{\em exponential speedup} if 
\begin{equation}
T_Q(L,\nu) = \mbox{Poly}(L)
\mbox{ for all $\nu$}.
\label{def:lqs}\end{equation}

According to this definition,
Shor's algorithm \cite{Shor} achieves exponential speedup
(if the factoring is not in BPP).
On the other hand, 
{\em the Deutsch-Jozsa algorithm \cite{DJ} does not achieve
exponential speedup} because
Deutsch's problem {\em is} in BPP.

\subsection{Extra qubits and redefinition of local sites}

In a quantum computer, there can exist many qubits 
that are not directly related to 
quantum computational speedup.
For example, 
one can replace a classical circuit that assists 
a quantum computer with a quantum circuit.
Then, the size of the quantum computer becomes larger
than the original one.
It is clear that, 
in the enlarged quantum computer, 
only the original part is relevant to quantum computational speedup.

Therefore, 
{\em we allow looking only at a subsystem of a quantum computer}
in order to find out its relevant part, 
whose state (according to our conjecture)
would have $q=2$.

Furthermore, 
one can add extra qubits and circuits to
a quantum computer 
without increasing $T_Q(L,\nu)$ by more than a Poly$(L)$ factor.
For example, 
to implement quantum error correction \cite{ShorCode,ftc}
one can replace each qubit with a logical 
qubit,  which is composed of $n$ qubits, 
where, e.g., $n=9$ for the Shor code \cite{ShorCode}.
In such a case, the correlation between two qubits 
is turned into 
a correlation between two {\em logical} qubits, 
i.e., a correlation among $2n $ qubits.
As a result, 
a state with $q=2$ of the original computer may change into
another state with $q<2$.
However, such a nonessential decrease in $q$ may be
recovered by regarding each {\em logical} qubit as a 
`local site'. 

Generally, in systems composed of discrete sites,
a local site (which may be, say, a quantum dot) physically has a finite spatial dimension.
A set of several neighboring sites 
also has a finite dimension.
Hence, 
the definition of a `local site' is to a great extent arbitrary.
It is therefore possible and reasonable to redefine a set of 
neighboring sites as a new single site \cite{2p-corr}.

In quantum computers, there is further arbitrariness because
it is possible to swap the states of two distant qubits, 
paying only a polynomial overhead.

From these observations, 
{\em we allow all possible redefinitions of local sites}
(accordingly, the number of local sites changes)
by regarding two or more qubits, however distant they are, as a local site.

\subsection{Generalized conjecture}\label{subsec:conj}

In order to apply 
the conjecture of refs.~\citen{conjecture1} and \citen{conjecture2}
to a large class of algorithms and models, 
we generalize it as follows.


{\em For a decision problem which is not in BPP, consider 
a quantum computer solving it. 
If the quantum computer achieves exponential speedup, 
then states with $q=2$, whose size is $\Omega(L)$, 
appear during computation, 
for some set $H$ of infinitely many instances;}
\begin{equation}
H \equiv \{ i(L,\nu) \ | \mbox{ some infinitely many $(L, \nu)$'s}  \}
\label{eq:instances.in.H}\end{equation}
{\em if `local sites' of the quantum computer are appropriately defined.}

This generalized conjecture can be rephrased as follows.
{\em 
After defining local sites appropriately, 
look at a certain subsystem composed of $\Omega(L)$ 
local sites of
the quantum computer.
Let $\hat \rho_k(L,\nu)$ be the reduced density operator of such a subsystem
in the $k$-th step of the quantum computation.
Take some function $k_*(L,\nu)$, which takes positive integral values, 
of $L$ and $\nu$.
For some set $H$ of infinitely many instances 
(eq.~(\ref{eq:instances.in.H})),  
consider the following family of states;
\begin{eqnarray}
&& \hspace{-14mm} F_{k_*}(H) 
\nonumber\\
&& \hspace{-12mm} \equiv \{ 
\hat \rho_{k_*}(L,\nu) \ | \
\mbox{all $\nu$ and $L$ such that } i(L,\nu) \in H
\}.
\label{eq:F}
\end{eqnarray}
If the quantum computer achieves exponential speedup,
one can find 
an appropriate definition of local sites, 
the function $k_*(L,\nu)$, 
and the set $H$, 
such that 
$q=2$ for the family $F_{k_*}(H)$.
}

We will explain the physical meaning of the set $H$
in the next subsection.

\subsection{Physical meaning of the set $H$}

If the above conjecture is correct, 
we can show that $H$
contains 
some infinitely many instances 
that cannot be solved within polynomial time by any classical computer.
That is, {\em $H$ 
contains infinitely many `hard' instances}.
This can be seen using reduction to absurdity as follows.

Suppose that the conjecture is correct but $H$
did {\em not} contain infinitely many instances that 
satisfy inequality (\ref{eq:TC>P}).
Then, there would exist a classical 
computer that solves all instances in $H$
within the time Poly$(L)$. By attaching this classical computer to $Q$ as a 
preprocessor, one could obtain another fast quantum computer, $Q'$.  
However, states with $q=2$ would not appear in $Q'$ at all, 
in contradiction to the conjecture. 
Therefore, 
$H$ must contain
infinitely many instances that satisfy inequality (\ref{eq:TC>P})
if our conjecture is correct.

Note that $H$ is {\em not} uniquely determined for a given problem
because an appropriate subset of $H$ can be another $H$.
To confirm the above conjecture, it is sufficient to 
find {\em one} of many possible $H$'s.

The set $H$ is closely related to a 
`complexity core' \cite{cc:Lynch,cc:ESY,cc:OS,cc:BD}. 
A {\em complexity core} 
(or polynomial complexity core) 
${\cal C}$ was defined by Lynch \cite{cc:Lynch} as
an infinite collection of instances such that every algorithm 
solving the problem using a deterministic Turing machine 
needs more than polynomial time almost everywhere on ${\cal C}$.
[Note that 
${\cal C}$ is not 
uniquely determined for a given problem
because an appropriate subset of ${\cal C}$ is also a
complexity core \cite{cc:Lynch,cc:ESY,cc:OS,cc:BD}.]
His idea has been generalized to complexity classes other than P
in refs.~\citen{cc:ESY,cc:OS,cc:BD}.
Intuitively, a complexity core is a set of `hard' instances.
The above-mentioned fact shows that {\em $H$ includes  
a BPP complexity core as a subset}.
Hence, our conjecture claims roughly that 
{\em states with $q=2$ appear for infinitely many instances in a 
complex core}.

\subsection{Remarks on the conjecture}

Before going further, we make a few remarks.

The conjecture 
does {\em not} claim that states with $q=2$ would be {\em sufficient}
for exponential speedup; 
it rather claims that they are {\em necessary}.
Hence, if states with $q=2$ appear in some quantum 
algorithm, it does not necessarily mean that 
the algorithm achieves exponential speedup.

Moreover, 
even when a quantum computer does achieve exponential speedup, 
the conjecture does {\em not} claim that 
{\em all} states with $q=2$ appearing in 
the computer would be relevant to exponential speedup.
In fact, we already showed in ref.~\citen{US04} that, 
although the final state $|\psi_{\rm DFT}\ket$ of the 
computation (before the final measurement) has $p=2$ (hence $q=2$),
it is irrelevant to exponential speedup.

Furthermore, 
for some states with $q=2$ of size $L$ (such as the GHZ state), 
one can construct  a quantum circuit that converts a product state into 
such a state only in $\Theta (L)$ steps 
{\em if the target state with $q=2$ is known beforehand} 
(i.e., when one constructs the circuit).
This has {\em nothing} to do with our conjecture.
The point is that,  
in quantum computation,  
the state that appears 
at, e.g., the middle point of the computation
is {\em unknown} when the circuit is constructed,
because the state varies considerably according to the instance.

Finally, we note that 
the results in ref.~\citen{US04} can be understood more clearly according to
the present generalized conjecture.
For example, it was shown in  ref.~\citen{US04} that 
states with $p=2$ do not appear when $r =2^n$,
where $n=1,2,3,\cdots$ 
and $r$ is the least positive integer that satisfies 
${\sf x}^{r}\equiv 1$ ($\mathrm{mod}\ {\sf N}$). 
[${\sf N}$ is a positive integer to be factored, 
and ${\sf x}$ a random integer co-prime to ${\sf N}$ 
that satisfies $0<{\sf x}<{\sf N}$ \cite{US04,Shor,EJ}.]
This fact does {\em not} conflict with the conjecture
because
such rare instances  do not affect the generalized index $p$ 
of \S\ref{sec:p}, 
which is defined not by $\Theta$ but by $\overline{\Theta}$.
It is also possible to exclude the instances with $r =2^n$ from $H$.

\subsection{Further extension to Grover's quantum search algorithm}
\label{sec:conj-Grover}

Shor's factoring algorithm \cite{Shor} 
is a representative algorithm for structured problems \cite{NC}.
For this algorithm, 
ref.~\citen{US04} supports the above conjecture.

On the other hand, a representative algorithm for unstructured problems
is Grover's quantum search algorithm \cite{Grover,NC}.
Hence, it is tempting to examine the conjecture in Grover's algorithm.
However, 
we cannot apply the conjecture (even in the above generalized form) 
directly to Grover's algorithm
because eq.~(\ref{def:lqs}) is not satisfied, 
i.e., 
it does not achieve exponential speedup. 
Nevertheless, it is often argued that the quadratic speedup of 
Grover's algorithm is significant \cite{NC}.
Furthermore, 
Grover's algorithm is optimal, 
i.e., 
no quantum algorithm is faster than Grover's algorithm 
by more than a Poly$(L)$ factor in solving the search problem \cite{NC}.
It is therefore very interesting to examine the conjecture, if possible, 
for Grover's algorithm.
To make it possible, we here extend the conjecture further to 
Grover's algorithm.
Its correctness for Grover's algorithm will be proved in the next section.

Grover's search problem is 
the problem of finding a solution to the equation $f_L(x) = 1$
among $N=2^L$ possibilities, where $f_L(x)$ is a function, 
$
f_L: \{0,1\}^{L} \mapsto \{0,1\}.
$
Let $M$ be the number of solutions and 
$x_1, \cdots, x_M$ be the solutions. 
For each $L$, the solutions specify an instance.
That is, $(x_1, \cdots, x_M)$ corresponds to $\nu$, which labels instances
as $i(L,\nu)$.
According to conventions, we regard the number of oracle calls 
as the computational time.

In extending the conjecture, we note that 
the degree of quantum speedup depends on the magnitude of $M$ 
[see \S\ref{sec:spdup-Grover}].
We here consider the case where 
\begin{equation}
M=O(2^{mL}) \quad (0 < m <1),
\label{eq:M=large}\end{equation}
where $m$ is a constant, independent of $L$.
In this case, the degree of quantum speedup is similar
to the case of $M=1$ [see \S\ref{sec:spdup-Grover}],
and hence seems significant.

We extend the conjecture of \S\ref{subsec:conj} 
simply by replacing the condition 
{\em `if the quantum computer achieves exponential speedup'}
with the relaxed condition
{\em `if the quantum computer achieves  exponential speedup
or it solves the quantum search problem 
using Grover's algorithm in the case of eq.~(\ref{eq:M=large})}'.

\section{Analytic Results for Grover's Quantum Search Algorithm}
\label{sec:Grover-analytic}

In this section, we show that the extended conjecture of
\S\ref{sec:conj-Grover} is correct for Grover's algorithm.

\subsection{Notation}
\label{sec:notation_Grover}

We first introduce the notation.
We assume that $N (=2^L) \gg M$ (number of solutions); 
otherwise, classical computers could solve the problem quickly.

It seems evident that an index register, composed of $L$ qubits, 
is relevant to Grover's algorithm. 
We therefore look only at the index register, although 
additional quantum circuits may be present in real quantum computers,
as discussed in \S\ref{subsec:conj}. 
For each instance $i(L, \nu)$, where 
$\nu=(x_1, \cdots, x_M)$ (see \S\ref{sec:conj-Grover}),
we put
\begin{eqnarray}
| \alpha(L, \nu) \rangle &\equiv& 
{1 \over \sqrt{N-M}} \sum_{x \, (\neq x_1, \cdots, x_M)} | x \rangle,
\label{eq:alpha}\\
| \beta(L, \nu) \rangle &\equiv& 
{1 \over \sqrt{M}} \sum_{x \, (= x_1, \cdots, x_M)} | x \rangle.
\label{eq:beta}\end{eqnarray}
Then the state $| \psi_0(L) \rangle$ just after 
the first Hadamard transformation (HT)
(see \S\ref{sec:simulation}) 
is represented as \cite{NC}
\begin{eqnarray}
| \psi_0(L) \rangle 
&=&
| \rightarrow \rightarrow \cdots \rightarrow \rangle
\nonumber\\
&=& 
\cos {\theta \over 2} \ | \alpha(L, \nu) \rangle 
+ 
\sin {\theta \over 2} \ | \beta(L, \nu) \rangle,
\label{eq:psi0=}
\end{eqnarray}
where $| \rightarrow \rangle \equiv (|0 \rangle + |1 \rangle)/\sqrt{2}$,
and the angle $\theta$ is given by
\begin{equation}
\cos {\theta \over 2} =
\sqrt{N - M \over N}.
\end{equation}
Let $\hat O(L, \nu)$ be the oracle operator;
\begin{equation}
\hat O(L, \nu) |x \rangle 
=
\begin{cases}
- |x \rangle & (x= x_1, \cdots, x_M), \\
|x \rangle & (\mbox{otherwise}).
\end{cases}
\end{equation}
The Grover iteration
\begin{equation}
\hat G(L, \nu) = \left[ 
2 | \psi_0(L) \rangle \langle \psi_0(L)| - \hat I(L) \right] 
\hat O(L, \nu)
\end{equation}
performs the rotation by the angle $\theta$ in the direction 
$| \alpha(L, \nu) \rangle \to | \beta(L, \nu) \rangle$
in the two-dimensional subspace spanned by 
$| \alpha(L, \nu) \rangle$ and $| \beta(L, \nu) \rangle$.
The state $| \psi_k(L, \nu) \rangle$ 
after $k$ ($=0, 1, 2, \cdots$) iterations is therefore given by  \cite{NC}
\begin{eqnarray}
| \psi_k(L, \nu) \rangle 
&=&
\hat G(L, \nu)^k | \psi_0(L) \rangle
\nonumber\\
&=&
\cos \left( {2k + 1 \over 2} \theta \right) | \alpha(L, \nu) \rangle 
\nonumber\\
& & + 
\sin \left( {2k + 1 \over 2} \theta \right) | \beta(L, \nu) \rangle.
\label{eq:G^kpsiHT=}
\end{eqnarray}
Hence, by repeating the Grover iteration 
\begin{equation}
R(L) \equiv \left\lceil 
\left(\arccos \sqrt{M/N} \right) /\theta \right\rceil
\simeq \left\lceil {\pi \over 4} \sqrt{N \over M}  \right\rceil
\end{equation}
times, the state evolves into 
$
| \psi_{R(L)}(L, \nu) \rangle 
\simeq 
| \beta(L, \nu) \rangle 
$.
Here, $\lceil a \rceil$ denotes 
the smallest integer among those larger than or equal to $a$. 
By observing this state in the computational basis, 
one can find a solution to the search problem with the probability
$\gtrsim 1/2$.
The range of $k$ is thus 
\begin{equation}
0 \leq k \leq R.
\end{equation}

\subsection{$M$ dependence of degree of speedup}
\label{sec:spdup-Grover}

According to convention, we regard the number of oracle calls 
as the computational time.
Then, apart from ${\rm Poly}(L)$ factors, 
\begin{equation}
T_Q(L,\nu) = R(L) = \Theta\left( \sqrt{N/M} \right) 
\label{eq:Q-grover}\end{equation}
for all instances. 
In contrast,  
for classical computers, 
there exist infinitely many instances (as $L \to \infty$) such that
\begin{equation}
T_C(L,\nu)= \Theta(N/M).  
\label{eq:Qwst-grover}\end{equation}
The quadratic speedup of $T_Q(L,\nu)$ over $T_C(L,\nu)$ 
is significant when $M$ is sufficiently small 
(such as $M= {\rm Poly}(L)$). 
On the other hand, {\em no} quantum speedup is achieved 
when $M$ is too large such as $M= N/{\rm Poly}(L)$
because then classical computers can solve the problem efficiently.

To be specific, 
we limit ourselves to the case of eq.~(\ref{eq:M=large}) where
$M=O(2^{mL})$ [$m$ is independent of $L$ and $0 < m <1$],
because almost all interesting applications of Grover's algorithm
would belong to this case.
For example, this includes the case of $M= {\rm Poly}(L)$, 
whereas the uninteresting case of $M= N/{\rm Poly}(L)$ is excluded.

When $M=O(2^{mL})$, 
we find that 
\begin{equation}
T_C(L,\nu) = \Theta \left(2^{(1-m)L} \right)
\end{equation}
for some infinitely many instances, and
\begin{equation}
T_Q(L,\nu) = \Theta \left(\sqrt{2^{(1-m)L}} \right)
\end{equation}
for all instances.
This quadratic speedup 
seems to be as significant as that of $M=1$.

\subsection{Family of states to evaluate $q$ or $p$}

As discussed above, 
the degree of speedup 
depends on how the number of solutions $M$ behaves asymptotically as
a function of $L$.
For clarity, 
we treat 
different asymptotic forms of $M$ separately
when investigating our extended conjecture.

Suppose that we are given a functional form of $M$, 
which asymptotically 
satisfies inequality (\ref{eq:M=large}), 
such as 
$M=L^2$. 
Then, it will turn out that the set of all instances
is an appropriate choice of $H$ of \S\ref{subsec:conj};
\begin{equation}
H = \{ i(L,\nu) \ | \mbox{ all } L, \mbox{ all $\nu$} \}.
\label{eq:allinstances}\end{equation}
To construct a family of states $F_{k_*}(H)$, i.e., eq.~(\ref{eq:F}), 
we specify the number $k_*$ as follows.
Since Grover's algorithm simply repeats the Grover iteration
$R(L)$ times, it seems natural to take $k_* = \lceil R(L)/2 \rceil$.
More generally, it seems reasonable to take 
\begin{equation}
k_* = \lceil R(L)/s \rceil, 
\label{eq:k*.Grover}\end{equation}
where $s$ is a positive constant independent of $L$.
It will turn out that this choice of $k_*$ is indeed appropriate.
A family of states $F_{k_*}(H)$ is thus constructed
for a given $k_*$ as
\begin{eqnarray}
F_{k_*}(H)
=
\{ 
| \psi_{k_*}(L,\nu) \rangle 
\ |  \mbox{ all } L, \mbox{ all $\nu$}
\}.
\label{eq:F.Grover}
\end{eqnarray}
Since all states of this family are pure, we will evaluate 
the index $p$ rather than $q$.

\subsection{When $M=1$}\label{sec:M=1}

When $M=1$, 
eq.~(\ref{eq:alpha}) reduces to
$| \alpha(L,\nu) \rangle 
= | \psi_0(L) \rangle + O(1/\sqrt{N})
= | \psi_0(L) \rangle + O(1/2^{L/2})
$, 
where $O(1/\sqrt{N})$ denotes a vector of length $O(1/\sqrt{N})$.
On the other hand, eq.~(\ref{eq:beta}) reduces to
$| \beta(L,\nu) \rangle = | x_1 \rangle$.
Since $| \psi_0(L) \rangle$ and $| x_1 \rangle$ are product states,
we find that $p=1$ for (families composed, respectively, of)
$| \psi_0(L) \rangle$, $| \alpha(L,\nu) \rangle$
and $| \beta(L,\nu) \rangle$.
Hence, $p=1$ for (families of) the initial and final states.

For intermediate states $| \psi_k(L,\nu) \rangle$'s 
of interest, 
it is convenient to investigate 
$\hat M_x \equiv \sum_l \hat \sigma_x(l)$
(which corresponds to the $x$ component of the total magnetization of
magnetic substances).
Note here that, 
in order to show that $p=2$, 
it is sufficient to find {\em one} additive observable 
(which in this case is $\hat M_x$)
that fluctuates macroscopically.
From eq.~(\ref{eq:G^kpsiHT=}), we find
\begin{equation}
\langle \psi_k(L,\nu) | \hat M_x | \psi_k(L,\nu) \rangle 
=
\cos^2 \left( {2k + 1 \over 2} \theta \right) L + O(1),
\label{eq:pkMxpk}
\end{equation}
\begin{equation}
\langle \psi_k(L,\nu) | \hat M_x^2 | \psi_k(L,\nu) \rangle 
=
\cos^2 \left( {2k + 1 \over 2} \theta \right) L^2 + O(L).
\label{eq:pkMx2pk}
\end{equation}
Hence, 
\begin{eqnarray}
\langle \psi_k(L,\nu) | ( \Delta \hat M_x )^2 | \psi_k(L,\nu) \rangle 
&=& 
{1 \over 4} \sin^2 \left( (2k + 1) \theta \right) L^2
\nonumber\\
& & + O(L).
\label{eq:pkMx2pk-2}
\end{eqnarray}
For all states in the family $F_{k_*}(H)$ of eq.~(\ref{eq:F.Grover}), we thus find
\begin{equation}
\langle \psi_{k_*}(L,\nu) | ( \Delta \hat M_x )^2 | \psi_{k_*}(L,\nu) \rangle 
\simeq
{1 \over 4} \sin^2 \left( {\pi \over s} \right) L^2
+ O(L).
\label{eq:Mx2_Fk*}\end{equation}
Since $s$ is independent of $L$, the right-hand side is $\Theta(L^2)$, 
and thus $p=2$ for this family.

\subsection{When $M=$ Poly$(L)$}\label{sec:M=Poly(L)}

We now consider the case where $M \geq 2$.
In this case, unlike in the case of $M=1$, 
$| \beta(L,\nu) \rangle$ has $p=2$ for some instances.

For example, suppose that $M=2$ and the two solutions for
some instance $\nu$ are 
$x_0 \equiv 1010 \cdots 10$ and $x_1 \equiv 0101 \cdots 01$.
Then, $p=2$ for 
$| \beta(L,\nu) \rangle = (| x_0 \rangle + | x_1 \rangle)/ \sqrt{2}$
(i.e., for the family $\{ | \beta(L,\nu) \rangle \ | \ \mbox{all } L \}$),
because
$\langle \beta(L,\nu) | ( \Delta \hat M_z^{\rm st} )^2 | \beta(L,\nu) \rangle = \Omega(L^2)$.
Here, $\hat M_z^{\rm st} \equiv \sum_l (-1)^l \hat \sigma_z(l)$,
which corresponds to the $z$ component of the staggered magnetization 
of antiferromagnets.
On the other hand, 
if $x_0 \equiv 0000 \cdots 00$ and $x_1 \equiv 0000 \cdots 01$ 
for another instance $\nu'$, 
then $p=1$ for 
$| \beta(L,\nu') \rangle = (| x_0 \rangle + | x_1 \rangle)/ \sqrt{2}$.

Therefore, {\em when $M \geq 2$,
$p$ of the final state depends on the instance, i.e., 
on the nature of the solutions.
}
(On the other hand, it is clear that 
$p=1$ for the initial state.)

To compute $p$ of states in intermediate stages of computation
for $M>2$, 
we first consider the case where $M=$ Poly$(L)$.
In this case, 
eq.~(\ref{eq:pkMx2pk-2}) still holds 
for $| \psi_k(L,\nu) \rangle$ with $k \geq 1$, 
and thus the argument following eq.~(\ref{eq:pkMx2pk-2}) also holds.
Therefore, 
we again find that
$p=2$ for the family $F_{k_*}(H)$ of eq.~(\ref{eq:F.Grover}).

We can obtain the same conclusion when 
$M=O(2^{L^\kappa})$, where $\kappa$ is a constant independent of $L$
and $0<\kappa<1$.
Instead of showing this, we shall derive 
the same conclusion when $M$ is even larger in the next subsection.

\subsection{When $M=\Theta(2^{mL})$}\label{sec:M=2mL}

We now study the case where 
\begin{equation}
M=\Theta(2^{mL}) \quad (0 < m <1),
\end{equation}
where $m$ is independent of $L$.
This is the upper limit of $M$ that satisfies the 
condition given by eq.~(\ref{eq:M=large}).

As mentioned above, 
$p$ of the final state depends on the nature of the solutions
(whereas $p=1$ for the initial state).
Since this might not be trivial when $M$ is as large as 
$\Theta(2^{mL})$, we give an example
for $M=2^{L/2}=\sqrt{N}$.
Suppose that the solutions for some instance $\nu$
are as follows:
$x_j = j$ for $j=1, 2, \ldots, \frac{1}{2}\sqrt{N}$, 
whereas $x_j$'s for $j \geq \frac{1}{2}\sqrt{N}+1$
are multiples of
$2\sqrt{N}$ (less than $N$). Then 
\begin{eqnarray}
|\beta(L,\nu) \rangle 
&=&
\frac{1}{\sqrt{2M}}\sum_{y=1}^{\frac{1}{2}\sqrt{N}}|00\cdots
  0y\rangle
\nonumber \\
& & +
\frac{1}{\sqrt{2M}}\sum_{z=0}^{\frac{1}{2}\sqrt{N}-1}|z0\cdots
  00\rangle 
\nonumber \\
 &=&
\frac{1}{\sqrt{2}}
\big(
|00\cdots 0 \rightarrow \rightarrow \cdots \rightarrow \rangle
\nonumber \\
& & +
|\rightarrow  \rightarrow \cdots \rightarrow 00\cdots 0\rangle 
\big)
+O\left({1 \over \sqrt{N}} \right).
\qquad
\end{eqnarray}
Since  
$\langle \beta(L,\nu) |(\Delta \hat{M'}_x)^2|\beta(L,\nu) \rangle =O(L^2)$,
where 
$\hat{M'}_x \equiv \sum_{l=1}^{L/2}\hat{\sigma
}_x(l)+\sum_{l=L/2+1}^{L}\hat{1 }(l)$,
we find that $p=2$ for this state.
By contrast, 
if $x_j = j$ for all $j$ for another instance $\nu'$, 
then 
$ |\beta(L,\nu') \rangle 
= |00 \cdots \rightarrow \rightarrow \cdots \rightarrow \rangle$,
for which $p=1$.

To compute $p$ of states in intermediate stages of computation, 
we note that 
$| \alpha(L,\nu) \rangle = | \psi_0(L) \rangle + O(2^{-(1-m)L/2})$
when $M = \Theta(2^{mL})$. Hence, 
abbreviating 
$| \psi_k(L,\nu) \rangle, |\alpha(L,\nu) \rangle$
and $|\beta(L,\nu) \rangle$ 
to
$| \psi_k \rangle, |\alpha \rangle$ and $|\beta \rangle$, respectively,
we find
\begin{eqnarray}
\langle \psi_k |\hat{M}_x | \psi_k \rangle 
&=& 
\cos^2\left( \frac{2k+1}{2}\theta \right)
\langle \alpha |\hat{M}_x|\alpha \rangle
\nonumber\\
& &
+
\sin^2\left( \frac{2k+1}{2}\theta \right)
\langle \beta |\hat{M}_x|\beta \rangle
+O(1),
\qquad
\\
\langle \psi_k |\hat{M}^2_x | \psi_k \rangle 
&=&
\cos^2\left( \frac{2k+1}{2}\theta \right)\langle \alpha |\hat{M}^2_x|\alpha \rangle
\nonumber\\
& &
+
\sin^2\left( \frac{2k+1}{2}\theta \right)\langle \beta |\hat{M}^2_x|\beta \rangle +O(1),
\end{eqnarray}
which yield
\begin{eqnarray}
&&
\langle \psi _k|(\Delta \hat{M}_x)^2|\psi _k\rangle 
\nonumber\\
&&
=
\cos^2\left( \frac{2k+1}{2}\theta \right)\langle \alpha |(\Delta \hat{M}_x)^2|\alpha \rangle
\nonumber\\
&&
+
\sin^2\left( \frac{2k+1}{2}\theta \right)\langle \beta |(\Delta \hat{M}_x)^2|\beta
  \rangle 
\nonumber \\
&&+
\frac{1}{4}\sin ^2((2k+1)\theta )\big{(} \langle
  \alpha |\hat{M}_x|\alpha \rangle -\langle \beta |\hat{M}_x|\beta
  \rangle \big{)}^2+O(L).
\qquad
\end{eqnarray}
As shown in Appendix \ref{appendix_matsuzaki}, 
$\langle \alpha |\hat{M}_x|\alpha \rangle =L +O(1)$
and $\langle \beta |\hat{M}_x|\beta \rangle \leq K L$,
 where $K$ is independent of $L$ and $0<K<1$. 
Hence, for $k_*$ of eq.~(\ref{eq:k*.Grover}), we find 
$\langle \psi_{k_*}|(\Delta \hat{M}_x)^2|\psi _{k_*} \rangle =O(L^2)$.
Therefore, $p=2$ 
for 
the family $F_{k_*}(H)$ of eq.~(\ref{eq:F.Grover}).

We have thus proved that 
the extended conjecture in \S\ref{sec:conj-Grover}
is correct for Grover's algorithm.

\subsection{Intermediate values of $p$}

As the quantum computation proceeds (i.e., as $k$ increases), 
$p$ increases from $1$ for the initial state $|\psi_0(L) \rangle$
to $2$ for $|\psi_{k_*}(L,\nu) \rangle$.
In the transient steps,
$p$ takes intermediate values between $1$ and $2$.

Such intermediate values 
are also taken, e.g., by states of quantum many-body systems 
at critical points of continuous phase transitions,
where two-point correlation functions decay according to power laws
as functions of the distance between two points.
Hence, one might expect some universal properties 
of $p$, as critical exponents in continuous phase transitions have.

For states of quantum computers in transient steps, however, 
the intermediate values of $p$ are not universal; 
they depend on details such as the nature of the solutions.
Since we are not interested in such nonuniversal values of $p$
in this paper, 
we have focused on the universal result,
which is directly related to our conjecture, that 
$p=2$ for 
a family composed of $|\psi_{k_*}(L,\nu) \rangle$'s.

\section{Evolution of Quantum Correlations 
in Grover's Quantum Search Algorithm}
\label{sec:Grover-numerical}

As summarized in \S\ref{sec:p.and.q}, 
the index $p$ 
is calculated from two-point correlations of local operators.
For pure states, $p=2$ if they have 
two-point correlations of $\overline{\Theta}(1)$ between  
$\Theta(L^2)$ pairs of sites.
[As discussed in \S\ref{sec:Pofpq}, 
for pure states, such strong two-point correlations
imply a strong $L$-point correlation.]
As discussed in \S\ref{sec:VCM},
the existence of such correlations can be detected by
the asymptotic behavior (as $L \to \infty$) 
of the maximum eigenvalue $e_{\rm max}$ 
of the VCM, 
because, roughly speaking, $L e_{\rm max}$ is 
proportional to the number of pairs of sites whose
correlation is of $\overline{\Theta}(1)$ \cite{rough}.

Although $p$ is simpler and more convenient for stating the conjecture, 
$e_{\rm max}$ has more detailed information about 
two-point correlations.
[For example, a state with $e_{\rm max}=L$ has stronger 
two-point correlations than that with $e_{\rm max}=L/10$,
whereas both states have $p=2$.]
It is therefore interesting to study how 
$e_{\rm max}$ evolves, as the computation proceeds,
from a small value (corresponding to $p=1$) in the initial state
to larger values.
It describes how two-point correlations evolve
(until a strong $L$-point correlation develops for $p=2$, as
discussed in \S\ref{sec:Pofpq}).

In this section, 
to investigate the evolution of $e_{\rm max}$, 
we numerically simulate a quantum computer that performs 
Grover's algorithm.
In the simulation, we study more states than 
those studied in the previous section, 
where 
we have studied 
$| \psi_k(L,\nu) \rangle =\hat G^k(L,\nu) | \psi_0(L) \rangle$
($k = 0, 1, 2, \cdots, R(L)$).
In actual quantum computations, 
the Grover iteration 
$\hat G(L,\nu)$ 
may be realized, e.g., as a series of local 
and pair-wise operations \cite{NC}.
Hence, many intermediate states appear between 
the computational steps corresponding to 
$| \psi_k(L,\nu) \rangle$ and $| \psi_{k+1}(L,\nu) \rangle$.
Although it was sufficient to investigate $| \psi_k(L,\nu) \rangle$'s
to confirm the conjecture, 
we also study such intermediate states
to see more details.

We simulate two cases, $M=1$ and $M=2$, 
because these cases are most fundamental. 
The solution(s) $x_1$ (and $x_2$) is chosen randomly. We have 
confirmed that this random choice of a solution(s)
makes no significant difference in the 
results of the numerical simulations presented below.

\subsection{Formulation of simulation}\label{sec:simulation}

We explain our simulation in the case of $M=1$.
Simulation for $M=2$ has also been performed similarly.

Since $M=1$, we can simply take $\nu=x_1$.
As in the previous section, 
we consider the index register composed of $L$ qubits. 
The register is initially set to be in the product state
\begin{equation}
|\psi_{\rm init}(L) \rangle = |00 \cdots 0 \rangle.
\label{psi0_G}\end{equation}
Firstly, the HT is performed
by successive applications of the Hadamard gate on individual qubits,
and the quantum state evolves into
$|\psi_0(L) \rangle$ of eq.~(\ref{eq:psi0=}). 
Then we apply the Grover iteration 
$\hat{G}(L,\nu)$, 
which consists of two HTs, 
an oracle operation $\hat{O}(L,\nu)$, 
and a conditional phase shift $\hat{P}(L)$ \cite{NC}; 
\begin{equation}
\hat{P}(L) |x \rangle 
=
\begin{cases}
|00 \cdots 0 \rangle & (x=0), \\
-|x \rangle & (x>0).
\end{cases}
\end{equation}

Each HT requires $L$ operations of 
the Hadamard gate. 
The oracle $\hat{O}(L,\nu)$ 
requires 
its own workspace qubits and computational time.
However, since the oracle is not a proper part of Grover's algorithm,
we simulate the operation of $\hat{O}(L,\nu)$ as a one-step operation,
and its workspace is not included in the simulation.
The execution of $\hat{P}(L)$ 
requires $\Omega(L)$ pairwise unitary operations.
For simplicity, however, we simulate $\hat{P}(L)$ as 
a one-step operation. 
Hence, each Grover iteration is simulated by 
$2L+2$ steps of operations.

After the application of the Grover iterations $R(L)$ times,
the state $|\psi_0(L) \rangle$ evolves into 
\begin{equation}
\hat{G}^{R(L)}(L,\nu)|\psi_0(L) \rangle
=|\psi_{R(L)}(L,\nu)\rangle
\simeq | \beta(L,\nu) \rangle
= |x_1 \rangle.
\label{GRpsi0}\end{equation}
Finally, by observing this state 
one can obtain the solution
$x_1$ with a sufficiently high probability.
We do not simulate this final measurement process.
The total computational time (steps) $T_Q(L,\nu)$ in our simulation is thus
\begin{equation}
T_Q(L,\nu)=L+ (2L+2)R(L) = \Theta(L\sqrt{2^{L}})
\end{equation}
for all $\nu$.
This is larger than $T_Q(L,\nu)$ in \S\ref{sec:Grover-analytic}
only by a polynomial factor.
For each instance $\nu$, 
$T_Q(L,\nu)$ different states,
including $| \psi_k(L, \nu) \rangle$'s in \S\ref{sec:Grover-analytic}, 
appear during computation.

In our conjecture, 
we have allowed 
(i) looking only at a subsystem
and (ii) all possible redefinitions of local sites.
For the present model of a quantum computer, however, 
we have confirmed 
(from the following results and the results of the previous section)
that they are unnecessary.
That is, in the present model, 
we can confirm the extended conjecture 
by simply calculating the index $p$ 
of states of the index register. 

To find states with $p=2$, 
we calculate 
$e_{\rm max}(L,\nu)$ of \S\ref{sec:VCM}.
By plotting the $L$ dependence of $e_{\rm max}(L,\nu)$, 
we can determine $p_e$ through eq.~(\ref{def:p_e}).
When states with $p_e=2$ are found, 
they have $p=2$ 
because, as discussed in \S\ref{sec:VCM},  
if $p_e=2$ then $p=2$ (and vice versa).
We will also plot how 
$e_{\rm max}(L,\nu)$ (for fixed $L$) grows and decays as 
the quantum computation proceeds 
because it is instructive and interesting.

In defining the VCM of eq.~(\ref{def:VCM}), 
we take $\hat b_\alpha(l)=\hat \sigma_\alpha(l)$ 
(the Pauli operator on site $l$
and $\alpha=1,2,3$), i.e,
\begin{equation}
V_{l \alpha, l' \alpha'}(L,\nu)
\equiv
\langle \Delta \hat \sigma_\alpha(l) 
\Delta \hat \sigma_{\alpha'}(l') \rangle_{L \nu}.
\label{def:V-2}\end{equation}
In the following, 
for conciseness,
we will often denote 
$e_{\rm max}(L,\nu)$, $V_{l \alpha, l' \alpha'}(L,\nu)$, and so on
simply by $e_{\rm max}$, $V_{l \alpha, l' \alpha'}$, and so on.

\subsection{Results of simulation for $M=1$}\label{ss:nr_Grover}

\begin{figure}[htbp]
\begin{center}
\includegraphics[width=0.95\linewidth]{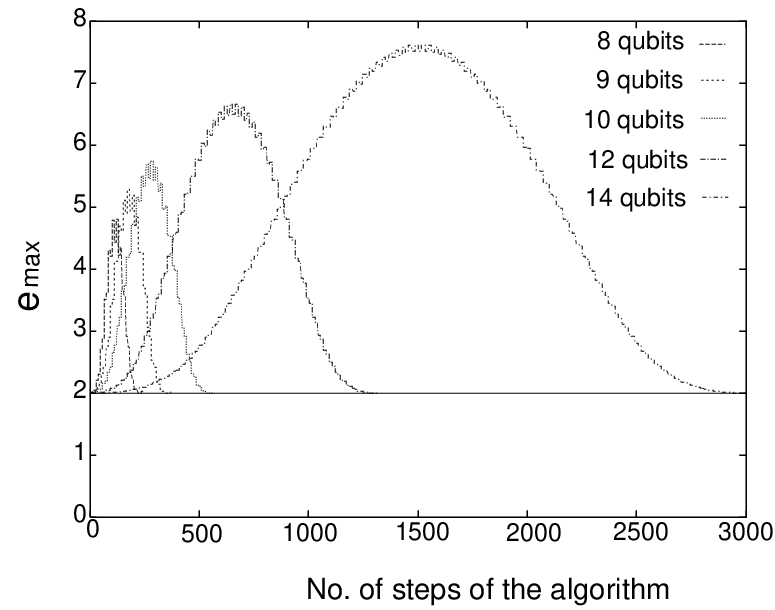}
\end{center}
\caption{
The maximum eigenvalue $e_{\rm max}$ of the VCM
of quantum states appearing in Grover's quantum search algorithm
for $L=8, 9, 10, 12$ and $14$
when $x_1 = 19, 388, 799, 1332$ and $9875$, respectively,
as functions of the number of step of the algorithm.
The horizontal line represents 
$e_{\rm max}$ for product states, 
$e_{\rm max}=2$.}
\label{Grover9}
\end{figure}

Figure \ref{Grover9} shows the evolution of 
$e_{\rm max}$ for $L=8, 9, 10, 12$ and $14$, 
when $x_1 = 19, 388, 799, 1332$ and $9875$, respectively,
for $M=1$.
The curves for different values of $L$ behave similarly in the sense 
that, on the whole, the curves are exponentially expanded 
along the horizontal axis as $L$ increases.

\begin{figure}[htbp]
\begin{center}
\includegraphics[width=0.95\linewidth]{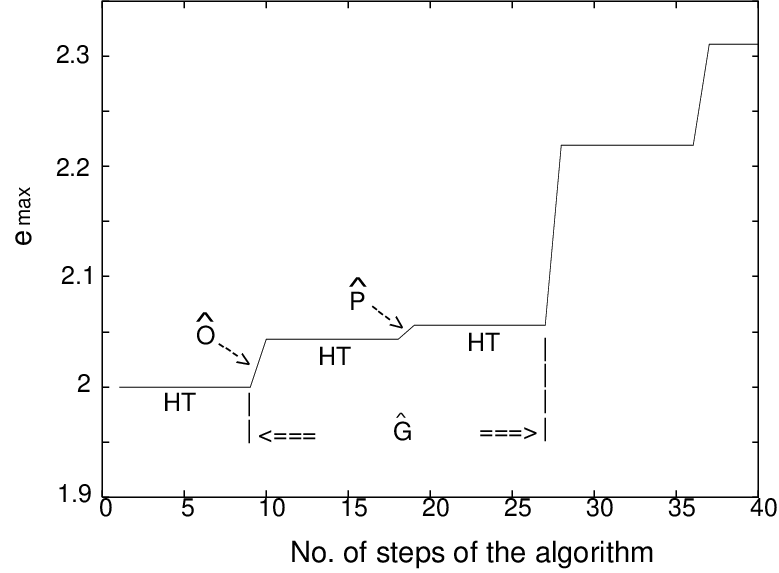}
\end{center}
\caption{
Magnification of Fig.~\ref{Grover9}, 
from the $1$st step to the $40$th step for $L=8$. 
Here, 
$\hat O$ and $\hat P$ represent
the oracle operation and conditional phase shift, respectively,
in a single Grover iteration  $\hat G$. 
}
\label{initial_G}
\end{figure}

\begin{figure}[htbp]
\begin{center}
\includegraphics[width=0.95\linewidth]{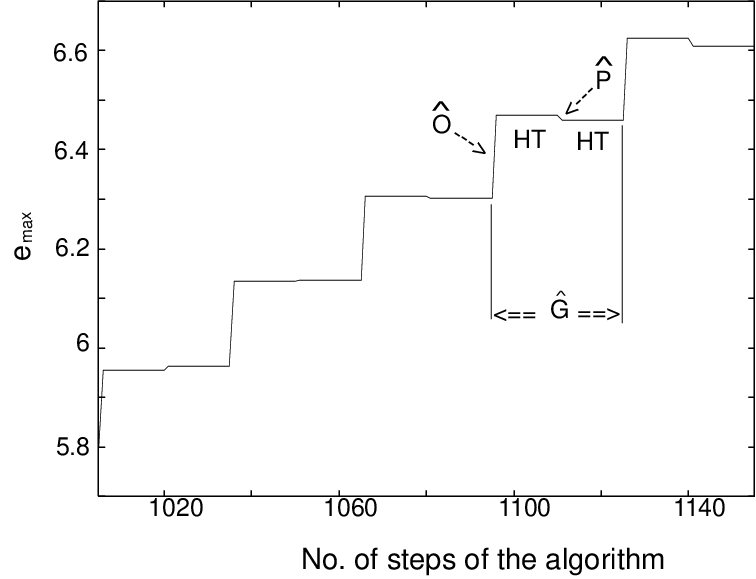}
\end{center}
\caption{
Magnification of Fig.~\ref{Grover9},
from the $1005$th step to the $1155$th step for $L=14$. 
Here, $\hat O$ and $\hat P$ represent
the oracle operation and conditional phase shift, respectively,
in a single Grover iteration  $\hat G$. 
}
\label{middle_G}
\end{figure}

Figure \ref{initial_G} is a magnification from the $1$st step 
to the $40$th step
for $L=8$, 
whereas Fig.~\ref{middle_G} shows a magnification from the $1005$th 
step to the $1155$th step for $L=14$.

It is seen that 
$e_{\rm max}=2.00$ for all states from 
$|\psi_{\rm init} \rangle$ to $|\psi_0 \rangle$, 
i.e., during the initial HT 
(from the $1$st step to the $8$th step in Fig.~\ref{initial_G}, 
denoted as `HT').
This is because all these states are product states, 
for which we can easily show that $e_{\rm max}=2$ 
(Appendix \ref{app:emax.product.state}). 
When the stage of Grover iterations begins, 
$e_{\rm max}$ grows gradually, as seen from 
Figs.~\ref{Grover9} and \ref{initial_G}.
In each Grover iteration, 
Figs.~\ref{initial_G} and \ref{middle_G} show that 
$e_{\rm max}$ changes when the oracle operator $\hat O$ is operated,
whereas it remains constant during the subsequent 
HT. 
Then, it changes again when $\hat P$ is operated,
whereas it remains constant again 
during the subsequent HT. 
As the Grover iterations are repeated, 
$e_{\rm max}$ (hence, quantum correlation) 
continues to increase as a whole, 
until it becomes maximum after about $R(L)/2$ 
applications of $\hat G$.
Further applications of $\hat G$ 
reduce $e_{\rm max}$, 
as seen from Fig.~\ref{Grover9},
toward $e_{\rm max} \simeq 2.00$ for $|\psi_{R(L)}\rangle$, 
which is approximately a product state as seen from eq.~(\ref{GRpsi0}).

\begin{figure}[htbp]
\begin{center}
\includegraphics[width=0.95\linewidth]{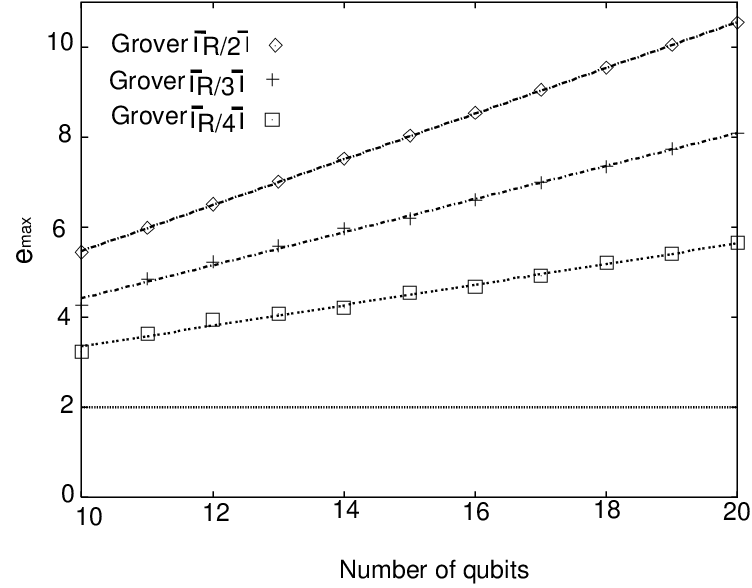}
\end{center}
\caption{
$e_{\rm max}$'s of $|\psi_{\lceil R/2 \rceil} \rangle$ (diamonds),
$|\psi_{\lceil R/3 \rceil} \rangle$  (crosses) 
and $|\psi_{\lceil R/4 \rceil} \rangle$ (squares)
as functions of $L$.
They all show that $p=p_e=2$.
The dotted lines are guides to the eyes,
whereas the horizontal line represents 
$e_{\rm max}$ for product states, $e_{\rm max}=2$.
}
\label{maxfluG}
\end{figure}

These results indicate that
our construction of the family of states, 
given by eqs.~(\ref{eq:k*.Grover}) and (\ref{eq:F.Grover}), 
is reasonable.
Although we have already shown in \S\ref{sec:Grover-analytic}
that $p=2$ for such a family, 
it is instructive to plot $e_{\rm max}(L,\nu)$ 
of $| \psi_{k_*}(L,\nu) \rangle$ as a function of $L$ 
when the solution $x_1$ is randomly chosen
(i.e., an instance $\nu$ is randomly chosen) for each $L$.
Figure \ref{maxfluG} shows 
$e_{\rm max}$'s of 
$|\psi_{\lceil R/2 \rceil} \rangle$,
$|\psi_{\lceil R/3 \rceil} \rangle$ and $|\psi_{\lceil R/4 \rceil} \rangle$
as functions of $L$
for such a randomly chosen $x_1$.
[We have confirmed that almost identical 
curves are obtained for other choices of $x_1$ as well.]
Since $e_{\rm max}$'s tend to be proportional to $L$ for large $L$, 
we can confirm that $p=p_e=2$.

\subsection{Results of simulation for $M=2$}\label{ss:nr_GroverM=2}

\begin{figure}[htbp]
\begin{center}
\includegraphics[width=0.95\linewidth]{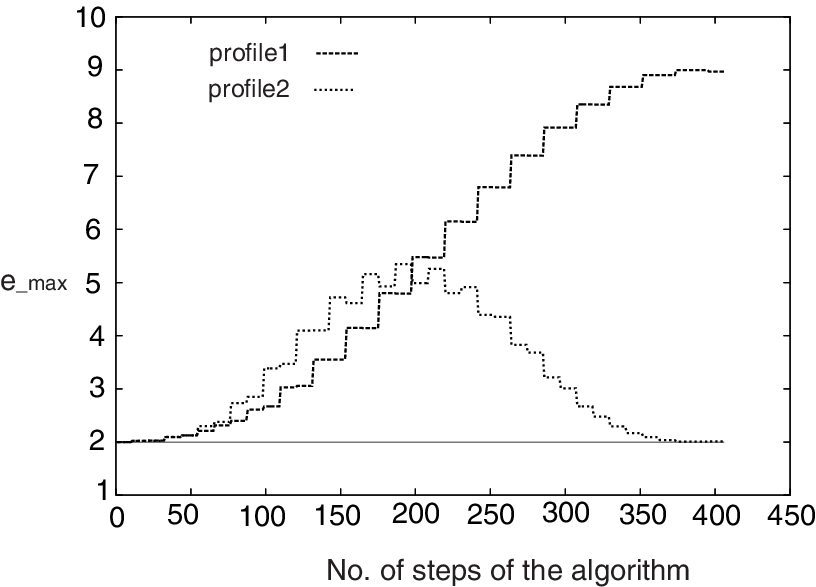}
\end{center}
\caption{
The maximum eigenvalue $e_{\rm max}$ of the VCM
of quantum states appearing in Grover's quantum search algorithm
for $L=10$ and $M=2$, 
when $x_1 = 2, x_2=1023$ (profile 1) and when 
$x_1 = 511, x_2=512$ (profile 2),
as functions of the number of steps of the algorithm.
The horizontal line represents 
the value of $e_{\rm max}$ for product states, $e_{\rm max}=2$.
}
\label{GroverM=2}
\end{figure}

When $M=2$, 
the result in \S\ref{sec:M=Poly(L)} indicates that 
$p=2$ for most states in the Grover iteration processes 
whereas $p$ of the {\em final} state depends on the nature of the solutions.
Figure \ref{GroverM=2} shows this clearly, where
we have plotted the evolution of 
$e_{\rm max}$ 
in two cases, i.e., $p=2$ (profile 1) and 
$p=1$ (profile 2) for the final state.
For 
$|\psi_{\lceil R/2 \rceil} \rangle$,
$|\psi_{\lceil R/3 \rceil} \rangle$ and $|\psi_{\lceil R/4 \rceil} \rangle$,
on the other hand, 
we obtain results similar to those in Fig.~\ref{maxfluG} in both cases.
Hence, $p=2$ for these states. 
This result visualizes how our extended conjecture holds when $M \geq 2$. 

\section{Discussion and Summary}\label{sec:dc}

We have studied the conjecture that 
the superposition of macroscopically distinct states is necessary for 
the significant speedup of quantum computers over classical computers.
This conjecture was previously supported for a circuit-based quantum computer
performing Shor's factoring algorithm.
To treat general algorithms and models, 
we have generalized the indices $p$ and $q$, which detect 
the superposition of macroscopically distinct states.
We then generalize the conjecture
in such a way that it is unambiguously applicable
to general models if a quantum algorithm achieves 
an exponential speedup.
We further extend the conjecture to the speedup 
achieved by Grover's quantum search algorithm.
This extended conjecture is proved to be correct for Grover's algorithm.
Since Grover's and Shor's algorithms are representative 
algorithms for unstructured and structured problems, 
respectively, 
the present results and the results of ref.~\citen{US04}
strongly support the conjecture.
To see details, 
we have also presented, by numerical simulation, 
how quantum correlation evolves and 
how the superposition of macroscopically distinct states
develops as the computation proceeds.

Jozsa and Linden previously showed that
entanglement over a cluster whose size is 
larger than $O(1)$ is necessary
for exponential speedup \cite{JL}.
For $p=2$, on the other hand, 
entanglement over a cluster whose size is 
larger than $O(L)$ is necessary \cite{SM02,MSS05}.
The present conjecture imposes a stronger condition in this sense.

Moreover, 
Vidal showed that a necessary condition for exponential 
speedup is that 
the amount of the bipartite entanglement 
between one part and the rest of the qubits
increases with $L$ \cite{Vidal}.
The bipartite entanglement studied by him 
is totally different from 
the entanglement associated with 
the superposition of macroscopically distinct states.
Therefore, 
the {\em simultaneous} requirement (for speedup over classical computations)
of Vidal's condition {\em and} 
the present conjecture 
is much stronger than the requirement of either one of them.
That is, for quantum speedup {\em both} 
the superposition of macroscopically distinct states
{\em and} 
sufficient bipartite entanglement 
are necessary.

It is also interesting to explore the relation 
between our results and the problem of 
time-optimal quantum evolution \cite{Carlini2006, Hosoya}.
In the latter case, the optimal evolution requires
a large fluctuation of the Hamiltonian operator, 
while in the former,  fast quantum 
computations require states with 
large fluctuations of additive operators.
This suggests a possible relation 
between quantum speedup and 
optimal evolution \cite{Hosoya}.

\begin{acknowledgments}

We thank M. Koashi and A. Hosoya
for valuable discussions.
This work was supported by 
PRESTO, Japan Science and Technology Corporation, 
by a Grant-in-Aid for Scientific Research No.\ 18-11581,
and by KAKENHI Nos.\ 22540407 and 23104707.

\end{acknowledgments}

\appendix

\section{$p=2$ implies superposition of macroscopically distinct states}
\label{app:p}

In this appendix, we explain why 
$p=2$ implies the superposition of macroscopically distinct states,
assuming for simplicity spatially homogeneous states.

Two states are macroscopically distinct if there is a macroscopic 
quantity whose expectation value is macroscopically distinct between 
them.
There are many quantities that are known as macroscopic quantities, 
including entropy, temperature, and pressure.
Among them, 
`additive mechanical variables', such as the total energy and the total magnetic moment, can be expressed by {\em additive operators}, 
as eq.~(\ref{def:A}).
[In contrast, `genuine thermodynamical variables', 
such as the entropy and temperature, 
cannot be expressed by additive operators.]

Let $L$ be the size of a macroscopic system.
In macroscopic physics such as thermodynamics, 
the values of 
additive mechanical variables scale as $\propto L$ with increasing $L$.
Therefore, the 
difference of an additive mechanical variable between two states 
is neglected if 
the difference is only $o(L)$ 
because the ratio of the difference to typical values vanishes in the macroscopic limit $L \to \infty$. 
In other words, 
two values of additive mechanical variables
are macroscopically distinct only when 
their difference scales as $\propto L$.
If one treats this macroscopic system by quantum theory, 
eigenvalues of an additive operator $\hat{A}(L)$ 
scale as $\propto L$.
Therefore, two eigenvalues or expectation values are macroscopically distinct 
only when their difference scales as $\propto L$.

Consider a state $|\psi(L) \rangle$ in a 
family of states of various values of $L$, $\{ |\psi(L) \rangle \}_L$. 
Let $| A, \mu ; L \rangle$ be an eigenvector
of $\hat{A}(L)$ corresponding to an eigenvalue $A$, 
where $\mu$ labels degenerate eigenvectors.
If $|\psi(L) \rangle$ is {\em not} a superposition of states 
with macroscopically distinct values of $A$, 
i.e., 
if it is just a superposition of $| A, \mu ; L \rangle$'s 
with macroscopically {\em non}distinct values of $A$ 
($+$ terms that vanish as $L \to \infty$),
then $|\langle A, \mu ; L |\psi(L) \rangle|^2$ takes nonvanishing
(or significant) values 
only for $A$ such that 
$\left| A - \langle \psi(L) | \hat{A}(L) |\psi(L) \rangle \right| = o(L)$.
In this case, 
$\langle \psi(L) | 
\Delta \hat{A}(L)^\dagger \Delta \hat{A}(L) 
|\psi(L) \rangle = o(L^2)$.
Hence, by contradiction, 
if 
$\langle \psi(L) | 
\Delta \hat{A}(L)^\dagger \Delta \hat{A}(L) 
|\psi(L) \rangle \neq o(L^2)$
then 
$|\psi(L) \rangle$ is a superposition of states 
with macroscopically distinct values of $A$.
Therefore, a {\em sufficient} condition for 
$|\psi(L) \rangle$ being a superposition of macroscopically distinct states
is that there exists an additive operator(s) $\hat{A}(L)$ 
(among many additive operators) such that
$\langle \psi(L) | 
\Delta \hat{A}(L)^\dagger \Delta \hat{A}(L) 
|\psi(L) \rangle = \Theta(L^2)$.
[The asymptotic notation such as $\Theta$ is summarized in Appendix \ref{app:notation}.]
This condition is simply expressed as
\begin{equation}
\max_{\hat{A}(L)} 
\langle \psi(L) | 
\Delta \hat{A}(L)^\dagger \Delta \hat{A}(L) 
|\psi(L) \rangle = \Theta(L^2).
\end{equation}
By definition, this means that $\{ |\psi(L) \rangle \}_L$ has $p=2$.
Therefore, $p=2$ implies the superposition of macroscopically distinct states.

For the argument in the present paper, 
we use this sufficient condition.
A state with $p<2$ may or may not be a superposition of 
macroscopically distinct states, depending on one's interest.
However, such states with $p<2$ are irrelevant to conclusions 
of the present paper.

Note that, in the above argument, we have {\em never} assumed that 
only {\em two} macroscopically distinct states are superposed to form 
$|\psi(L) \rangle$ with $p=2$.
Therefore, 
$|\psi(L) \rangle$ with $p=2$ is 
a superposition of {\em two or more} macroscopically distinct states.

To illustrate how $p$ is useful, we give a few simple examples.

For the GHZ or `cat' state,
$
| \psi_{\rm cat}(L) \rangle 
\equiv
(
|00 \cdots 0 \rangle
+
|11 \cdots 1 \rangle
)/\sqrt{2}
$,
a heuristic discussion would be sufficient to determine that 
it is a superposition of macroscopically distinct states,
because it is simply 
a superposition
of {\em two} macroscopically distinct states
and their coefficients do {\em not} vanish as $L \to \infty$.

On the other hand, 
let us consider the following state;
\begin{eqnarray}
&&
{1 \over \sqrt{L-1}}
\big(
|100 \cdots 00 \rangle
+
|110 \cdots 00 \rangle
+
\cdots
+
|111 \cdots 10 \rangle
\big)
\nonumber\\ 
&& \quad =
{1 \over \sqrt{L-1}}
\sum_{j=1}^{L-1}
| j ; j+1 \rangle,
\label{eq:dw.2}
\end{eqnarray}
where 
$| j ; j+1 \rangle$
is the state that has a `domain wall' between sites $j$ and $j+1$.
In this sum, 
each state differs from the preceeding state {\em only in a single qubit},
and the weight of each state {\em vanishes} in the $L \to 0$ limit.
For such a state, heuristic discussions will be ambiguous.
Even for such a state, by using the index $p$, 
we can easily show that it is a superposition of
macroscopically distinct states because 
it has $p=2$.

By contrast, 
the `W state'
$
\left(
|100 \cdots 0 \rangle
+
|010 \cdots 0 \rangle
+
\cdots
+
|000 \cdots 1 \rangle
\right)
/\sqrt{L}
$
has $p=1$,
hence one {\em cannot} say that it is 
a superposition of macroscopically distinct states \cite{SM02,MSS05}.
This is reasonable because
the W state corresponds to 
normal states in condensed-matter physics, 
such as a Frenkel exciton excited in an insulating solid \cite{US04},
which can be created easily in experiments.
In contrast, states with $p=2$ are extremely abnormal
in view of many-body physics,
as discussed in \S\ref{sec:Pofpq} and ref.~\citen{SM02}. 

Regarding the index $q$ for mixed states, see ref.~\citen{SM05}
for its physical meaning and several examples.

\section{On the restriction that $\| \hat a(l) \|=1$}
\label{app:why|a|=1}

Suppose that 
$\hat A = \sum_{l} \hat a(l)$
is an additive operator.
In the present paper, we have required that $\| \hat a(l) \|=1$.
To understand technical details about this, 
the following examples would be helpful.
\begin{itemize}
\item[ex.1]
The operator
$\hat A''_1 \equiv \sum_{l} \hat a(l)/2$
is {\em not} an additive operator according to 
the present definition, because the norm of the 
local operators is not unity.
However, by simply multiplying $\hat A''_1$  by $2$,  
we can 
obtain an additive 
operator $\hat A_1 \equiv 2\hat A''_1$ ($=\hat A$).
The fluctuations of 
$\hat A_1$ and $\hat A''_1$ differ only by a constant factor.

\item[ex.2] 
The operator
$\hat A''_2 \equiv \sum_{l = {\rm odd}} \hat a(l)$ 
is {\em not} an additive operator according to 
the present definition, because the norm of the 
local operators for even $l$ vanishes.
However, 
this operator has the same fluctuation as the additive operator
$\hat A_2
\equiv \sum_{l = {\rm odd}} \hat a(l) + \sum_{l = {\rm even}} \hat 1(l)$.

\item[ex.3] 
The operator
$\hat A''_3 \equiv \sum_l  \left(1 + {(-1)^l \over 2} \right) \hat a(l)$
is {\em not} an additive operator according to 
the present definition. 
However, there always exist real numbers $\alpha, \beta(l)$ such that 
$\hat a_3(l) \equiv \alpha \left(1 + {(-1)^l \over 2} \right) \hat a(l)
+ \beta(l) \hat 1(l)$ 
has a unit norm for every $l$.
Then, $\hat A_3 \equiv \sum_l  \hat a_3(l)$ is an 
additive operator, and 
the fluctuations of $\hat A_3$ and $\hat A''_3$
differ only by a constant factor.
\end{itemize}
Therefore, operators like $\hat A''_1, \hat A''_2$ and $\hat A''_3$ 
are essentially included 
(as $\hat A_1, \hat A_2$ and $\hat A_3$, respectively) 
when 
taking $\max_{\hat A(L)}$ in eq.~(\ref{def:p}).

The point is that 
one can modify $\hat A''_1, \hat A''_2$ and $\hat A''_3$
in such a way that the fluctuations of 
the modified operators 
$\hat A_1, \hat A_2$ and $\hat A_3$
(which are additive operators),
respectively, 
have the same order of magnitude 
as those of $\hat A''_1, \hat A''_2$ and $\hat A''_3$.
Note that this modification is {\em not} unique. For example, 
from $\hat A''_1$, one can also construct the additive operator
$2 \hat A''_1 - \hat a(1) +\hat 1 (1) 
= \hat 1(1) + \sum_{l \geq 2} \hat a(l)$,
which has the same order of fluctuation as $\hat A_1$.
This nonuniqueness does not cause any difficulty 
because $p$ and $q$ are defined by the order of magnitude
of fluctuations.

\section{Asymptotic notation}\label{app:notation}

Let $f(L)$ and $g(L)$ be non-negative functions of a positive variable $L$.
Following ref.~\citen{NC}, we use the following asymptotic notation:
\begin{eqnarray}
f(L) = O(g(L)) &\Leftrightarrow&
f(L)  \leq K g(L),
\\
f(L) = \Omega(g(L)) &\Leftrightarrow&
J g(L) \leq f(L),
\\
f(L) = \Theta(g(L)) &\Leftrightarrow&
J g(L) \leq f(L) \leq K g(L),
\end{eqnarray}
as $L \to \infty$, where $J$ and $K$ are some positive constants.

Let $f_\nu(L)$'s be non-negative functions,
which are labeled by an index $\nu$,
of a positive variable $L$.
We consider a family that consists of these functions, 
i.e., a family of real values, $\{ f_\nu(L) \}_{L,\nu}$.
Assuming that the number of possible values of $\nu$ 
increases to infinity as $L \to \infty$,
we use the following asymptotic notation:
\begin{eqnarray}
f_\nu(L) = \overline{O}(g(L)) &\Leftrightarrow&
f_\nu(L) = O(g(L))
\nonumber\\
&& \quad
\mbox{ for almost every $\nu$},
\\
f_\nu(L) = \overline{\Omega}(g(L)) &\Leftrightarrow&
f_\nu(L) = \Omega(g(L))
\nonumber\\
&& \quad
\mbox{ for almost every $\nu$},
\\
f_\nu(L) = \overline{\Theta}(g(L)) &\Leftrightarrow&
f_\nu(L) = \Theta(g(L))
\nonumber\\
&& \quad
\mbox{ for almost every $\nu$}.
\end{eqnarray}

For example, 
if $\nu = 1, 2, \ldots, 2^L$ for each $L$ and 
$f_\nu(L)=L^3 (1+\sin L) (1/L-1/L^{\nu})$,
then $f_\nu(L)=\overline{\Theta}(L^2)$, 
whereas it is {\em not} $\Theta(L^2)$.

\section{Range of $p$}\label{appendix_range}

In this appendix, we show that $1 \leq p \leq 2$.
As the operator norm $\| \hat a \|$,
we employ $\| \hat a \|_E$, defined by eq.~(\ref{norm.E}), 
in this Appendix.
Since
$ 
\left\langle \left( \Delta \hat A(L) \right)^2 \right\rangle_{L \nu}
\leq
\left\langle \hat A(L)^2 \right\rangle_{L \nu}
\leq
L^2,
$ 
we find that $p \leq 2$.
To prove that $p \geq 1$, we use the following Lemma;\\
{\em Lemma:}
For any state, which can be a mixed state,
there always exists a local operator $\hat a_*(l)$ that satisfies
\begin{equation}
\| \hat a_*(l) \|_E=1 \ \mbox{and} \ 
\langle \Delta \hat a_*(l)^\dagger\Delta \hat a_*(l) \rangle=1,
\end{equation}
{\em Proof:}
For a given state $\hat \rho(L)$, 
its local density operator $\hat \rho_l \equiv
{\rm Tr}_{l' (\neq l)} \hat \rho(L)$
can be diagonalized as
\begin{equation}
\hat \rho_l = \sum_{j=1}^d w_j(l) |j,l \rangle \langle j, l|,
\quad \sum_{j=1}^d w_j(l)=1,
\end{equation}
where $\{ |j,l \rangle \}_j$ is a complete orthonormal set of site $l$.
Take
\begin{equation}
\hat a_*(l) = \sum_{j={\rm odd}} 
\left( |j,l \rangle \langle j+1, l| + h.c. \right)
\end{equation}
where $\langle d+1, l| \equiv \langle 1, l|$.
This is a Hermitian local operator with 
$\| \hat a_*(l) \|_E=1$.
Since 
${\rm Tr} \left[ \hat \rho (L) \hat a_*(l) \right] =0$
and
\begin{equation}
\hat a_*(l)^\dagger \hat a_*(l)
=\sum_{j=1}^d |j,l \rangle \langle j, l|=\hat 1(l),
\end{equation}
we find that 
$
\langle \Delta \hat a_*(l)^\dagger\Delta \hat a_*(l) \rangle
=1
$.
$\blacksquare$

Using this Lemma, we now show the following theorem, from which 
it is evident that $p \geq 1$.\\
{\em Theorem:}
For any state, which can be a mixed state,
there always exists an additive operator that satisfies
\begin{equation}
\langle \Delta \hat{A}(L)^\dagger \Delta \hat{A}(L)\rangle \geq L.
\end{equation}
{\em Proof:}
We use the induction method.
We define
\begin{equation}
\hat A(k) \equiv \sum_{l=1}^k \hat a(l)
\quad (\| \hat a(l) \| = 1),
\end{equation}
for $1 \leq k \leq L$.
When $k=1$, the above Lemma shows that 
there exists $\hat{A}(1)$ such that 
$\langle \Delta \hat{A}(1)^\dagger \Delta \hat{A}(1)\rangle \geq 1$.
Now, assume that 
there exists $\hat{A}(k)$ such that 
$\langle \Delta \hat{A}(k)^\dagger\Delta \hat{A}(k)\rangle \geq 1$.
From the above Lemma, 
there exists a local operator $\hat{a}_*(k+1)$ on site $k+1$ such that 
$\| \hat{a}_*(k+1)\|=1$ and 
$
\langle \Delta \hat{a}_*(k+1)^\dagger \Delta \hat{a}_*(k+1)\rangle =1.
$
So, construct $\hat{A}(k+1)$ as
\begin{equation}
\hat{A}(k+1) = \hat{A}(k)+\hat{a}_*(k+1).
\end{equation}
Then,
\begin{eqnarray}
&&
\langle \Delta \hat{A}(k+1)^\dagger \Delta \hat{A}(k+1)\rangle
\nonumber\\
&& \quad
=
\langle \Delta \hat{A}(k)^\dagger\Delta \hat{A}(k)\rangle 
\nonumber\\
&& \qquad + 
\left\{ 
\langle \Delta \hat{A}(k)^\dagger \Delta \hat{a}_*(k+1) \rangle
+ c.c. \right\}
+1.
\end{eqnarray}
Therefore,
if 
$
\left\{ 
\langle \Delta \hat{A}(k)^\dagger \Delta \hat{a}_*(k+1) \rangle
+ c.c. \right\} \geq 0
$,
then
$\langle \Delta \hat{A}(k+1)^\dagger \Delta \hat{A}(k+1)\rangle \geq k+1$.
If, on the other hand, 
$
\left\{ 
  \langle \Delta \hat{A}(k)^\dagger \Delta \hat{a}_*(k+1) \rangle
+ c.c. \right\} <0
$,
then reconstruct $\hat{A}(k+1)$ as
\begin{equation}
\hat{A}(k+1) = \hat{A}(k)-\hat{a}_*(k+1).
\end{equation}
This gives 
$\langle \Delta \hat{A}(k+1)^\dagger \Delta \hat{A}(k+1)\rangle \geq k+1$.
$\blacksquare$

\section{Approach of a pure state with $p=2$ to a mixed state 
as $L \to \infty$}
\label{app:pure2mixed}

It was proved rigorously that 
any {\em pure} state with $p=2$ in a {\em finite} system of size $L$ 
does {\em not} approach pure states in an {\em infinite} 
system as $L \to \infty$ \cite{SM02}.
Although this might sound strange to the reader who is not familiar with 
the quantum theory of infinite systems \cite{Haag}, 
its physics can be understood as follows.

In the quantum theory of finite systems, 
all possible representations are equivalent (unitary equivalence).
In the quantum theory of infinite systems, by contrast, 
many representations that are {\em not} equivalent
to each other can exist \cite{Haag}.
Among two or more inequivalent representations, we have to choose
one that is suitable for describing the physical states 
of interest \cite{Haag}.
This makes the quantum theory of infinite systems 
very different from that of 
finite systems.
The above strange fact comes from this great difference.

As the simplest example, 
consider a cat state 
$
| \psi(L) \rangle 
\equiv
(
|00 \cdots 0 \rangle
+
|11 \cdots 1 \rangle
)/\sqrt{2}
$ 
of  size $L$.
If $L$ is finite, there exist observables that have nonvanishing 
matrix elements between $|00 \cdots 0 \rangle$ and 
$|11 \cdots 1 \rangle$.
The expectation values of such observables
discriminate between the pure state $| \psi(L) \rangle$ 
and the mixed state
$
\rho(L)
\equiv
(
|00 \cdots 0 \rangle \langle 00 \cdots 0 |
+
|11 \cdots 1 \rangle \langle 11 \cdots 1 |
)/2
$.
If we take the $L \to \infty$ limit, however, 
the quantum theory of infinite systems requires that  
every observables should be a function of field operators 
within a {\em finite} region in an infinite space \cite{Haag}.
As a result, 
there are no observables that have nonvanishing 
matrix elements between the $L \to \infty$ limits of
$|00 \cdots 0 \rangle$ and $|11 \cdots 1 \rangle$.
This implies that 
$\lim_{L \to \infty} | \psi(L) \rangle$
is not a pure state.
Here, the rigorous definition of pure states in ref.~\citen{Haag} is used
instead of the (over)simplified definition $\hat{\rho}^2=\hat{\rho}$,
because the latter can be used only for (an irreducible representation 
for) finite systems.

More mathematically speaking, 
if $| \psi(L) \rangle$ has $p=2$ then
$\lim_{L \to \infty} | \psi(L) \rangle$
is not a vector state of an irreducible representation.
For details, see ref.~\citen{SM02} and references cited therein.

\section{Equivalence of $p_e=2$ and $p=2$}
\label{app:C=O(L^0)}

In this appendix, we show that 
if $p_e=2$ then $p=2$ and vice versa.
For simplicity, we will omit `$\nu$' and `$(L)$', i.e., 
we will abbreviate $\hat A'(L), | \psi_\nu(L) \rangle, C_\nu (L)$ 
and so on to $\hat A', | \psi \rangle, C$ and so on, respectively. 
We first note that 
\begin{eqnarray}
\left\| 
\Delta \hat A' | \psi \rangle
\right\|^2
&=&
\left\| 
\sum_{l=1}^L \sum_{\alpha =1}^D c'_{l \alpha} \Delta \hat b_\alpha(l)
| \psi \rangle
\right\|^2
\nonumber\\
&\leq&
\left[
\sum_{l=1}^L \sum_{\alpha =1}^D 
\left| c'_{l \alpha} \right|^2 
\right]
\left[
\sum_{l=1}^L \sum_{\alpha =1}^D 
\left\| 
\Delta \hat b_\alpha(l)| \psi \rangle
\right\|^2
\right]
\nonumber\\
&\leq&
L \overline{O}(L),
\end{eqnarray}
from which $p_e \leq 2$.
Here, we have used the inequality 
$
\sum_k \left| \sum_j x_j y_{jk} \right|^2
\leq
\sum_k \sum_j \left| x_j \right|^2 \sum_{j'} \left| y_{j'k} \right|^2
=
\sum_j \left| x_j \right|^2 \sum_{j'} \sum_k \left| y_{j'k} \right|^2
$,
which holds for arbitrary complex numbers $x_j, y_{jk}$.
On the other hand, it is clear from the definitions that 
\begin{equation}
p \leq p_e.
\label{p<p_e}\end{equation}
Therefore, if $p=2$ then $p_e=2$.

To show the inverse, we assume that $p_e=2$, i.e., 
eq.~(\ref{fluc:DAmax}) holds.
Without loss of generality, we also assume that 
$ 
\max_{l, \alpha} 
\left| c^{\prime}_{l \alpha \, {\rm max}} \right|^2 
=
\left| c^{\prime}_{1 1 \, {\rm max}} \right|^2.
$ 
It is clear from eq.~(\ref{normalization:c'max}) that
\begin{equation}
\left| c^{\prime}_{1 1 \, {\rm max}} \right|^2 
\leq L.
\label{eq:c11<L}
\end{equation}
Our purpose is to show that 
\begin{equation}
\left| c^{\prime}_{1 1 \, {\rm max}} \right|^2 
= \overline{O}(L^0),
\label{eq:c11=O(1)}\end{equation}
because it yields
$C = \overline{O}(L^0)$, which 
gives eq.~(\ref{fluc:DAmax:normalized}) (implying $p=2$).
Equation (\ref{fluc:DAmax}) can be rewritten as
\begin{eqnarray}
&&
\sum_{\alpha, \alpha' =1}^D
c^{\prime \, *}_{1 \alpha \, {\rm max}}
c^{\prime}_{1 \alpha' \, {\rm max}}
V_{1 \alpha, 1 \alpha'}
\nonumber\\
&& \quad +
\left(
\sum_{l=2}^L \sum_{\alpha, \alpha' =1}^D
c^{\prime \, *}_{l \alpha \, {\rm max}}
c^{\prime}_{1 \alpha' \, {\rm max}}
V_{l \alpha, 1 \alpha'}
+ c.c.
\right)
\nonumber\\
&& \quad +
\sum_{l,l'=2}^L \sum_{\alpha, \alpha' =1}^D
c^{\prime \, *}_{l \alpha \, {\rm max}}
c^{\prime}_{l' \alpha' \, {\rm max}}
V_{l \alpha, l' \alpha'}
=
\overline{O}(L^2).
\nonumber\\
\label{eq:DAmax_decomp}\end{eqnarray}
The first term is positive 
(because the VCM is a non-negative Hermitian matrix) 
and $\leq \overline{O}(L)$.
The second term is estimated as
\begin{eqnarray}
&&
\left|
\sum_{l=2}^L \sum_{\alpha, \alpha' =1}^D
c^{\prime \, *}_{l \alpha \, {\rm max}}
c^{\prime}_{1 \alpha' \, {\rm max}}
V_{l \alpha, 1 \alpha'}
\right|
\nonumber\\
&& \quad \leq
D |c'_{1 1 \, {\rm max}}|
\left|
\sum_{l=2}^L \sum_{\alpha =1}^D 
c^{\prime \, *}_{l \alpha \, {\rm max}}
\right|
\max_{l, \alpha, \alpha'} \left|
V_{l \alpha, 1 \alpha'}
\right|
\nonumber\\
&& \quad \leq
D |c'_{1 1 \, {\rm max}}|
\sqrt{D(L-1)} \sqrt{L} \max_{l, \alpha, \alpha'} \left|
V_{l \alpha, 1 \alpha'}
\right|
\nonumber\\
&& \quad \leq
|c'_{1 1 \, {\rm max}}| \overline{O}(L)
\leq \overline{O}(L^{3/2}),
\label{2ndterm}\end{eqnarray}
where we have used inequality (\ref{eq:c11<L}) and 
$
| \vec{x} \cdot \vec{1} |
\leq | \vec{x} | |\vec{1} | = n | \vec{x} |
$,
which holds for an arbitrary $n$-dimensional vector $\vec{x}$.
Therefore, the third term of eq.~(\ref{eq:DAmax_decomp}) should be
\begin{equation}
\sum_{l,l'=2}^L \sum_{\alpha, \alpha' =1}^D
c^{\prime \, *}_{l \alpha \, {\rm max}}
c^{\prime}_{l' \alpha' \, {\rm max}}
V_{l \alpha, l' \alpha'}
=
\overline{O}(L^2).
\label{3rdterm}\end{equation}
On the other hand, let 
\begin{equation}
c^{\prime}_{l \alpha}
\equiv
{1
\over
\sqrt{
1-
\sum_{\alpha=1}^D |c^{\prime}_{l \alpha \, {\rm max}}|^2 / L
}}
c^{\prime}_{l \alpha \, {\rm max}},
\end{equation}
then the following operator,
which does not involve an operator on site $1$, 
takes the form of eq.~(\ref{DA:sum=L}):
\begin{equation}
\Delta \hat{A}^{\prime}
\equiv
\sum_{l=2}^L \sum_{\alpha =1}^D 
c^{\prime}_{l \alpha}
\Delta \hat b_\alpha(l).
\label{DA''max}
\end{equation}
We therefore have, using eqs.~(\ref{2ndterm}) and (\ref{3rdterm}), 
\begin{eqnarray}
0 &\leq&
\langle \Delta \hat{A}^{\prime \nu \dagger}_{\rm max} 
\Delta \hat{A}^{\prime \nu}_{\rm max} \rangle_{L \nu}
-
\langle \Delta \hat{A}^{\prime \nu \dagger} 
\Delta \hat{A}^{\prime \nu} \rangle_{L \nu}
\nonumber\\
&\leq&
\sum_{\alpha, \alpha' =1}^D
c^{\prime \, *}_{1 \alpha \, {\rm max}}
c^{\prime}_{1 \alpha' \, {\rm max}}
V_{1 \alpha, 1 \alpha'}
\nonumber\\
&&+
\left(
\sum_{l=2}^L \sum_{\alpha, \alpha' =1}^D
c^{\prime \, *}_{l \alpha \, {\rm max}}
c^{\prime}_{1 \alpha' \, {\rm max}}
V_{l \alpha, 1 \alpha'}
%
%
+ c.c.
\right)
\nonumber\\
& & -
{|c'_{11 \, {\rm max}}|^2 \over L}
\sum_{l,l'=2}^L \sum_{\alpha, \alpha' =1}^D
c^{\prime \, *}_{l \alpha \, {\rm max}}
c^{\prime}_{l' \alpha' \, {\rm max}}
V_{l \alpha, l' \alpha'}
\nonumber\\
&\leq&
|c'_{11 \, {\rm max}}|^2 \overline{O}(L^0)
+
|c'_{11 \, {\rm max}}| \overline{O}(L)
-
{|c'_{11 \, {\rm max}}|^2 \over L} \overline{O}(L^2)
\nonumber\\
&\leq&
|c'_{11 \, {\rm max}}|
\left[
|c'_{11 \, {\rm max}}| \overline{O}(L^0)
+
\overline{O}(L)
-
|c'_{11 \, {\rm max}}|\overline{O}(L)
\right].
\nonumber\\
\end{eqnarray}
Therefore, eq.~(\ref{eq:c11=O(1)}) should be satisfied
because, otherwise, the last line would become negative in contradiction with 
the first line.

For a family of homogeneous states, in particular, 
$p=p_e$ for every value of $p$,
because in this case the VCM has the translational invariance, 
and thus an eigenvector 
corresponding to $e_{\rm max}$ is also translational-invariant.
[Even when $e_{\rm max}$ is a degenerate eigenvalue,
one can construct a translational-invariant eigenvector 
by taking a linear combination of eigenvectors corresponding to $e_{\rm max}$.]
From eq.~(\ref{normalization:c'max}), 
this means $C = \overline{O}(L^0)$ for the additive operator 
that is composed of such an eigenvector, 
and thus $p \geq p_e$.
From inequality (\ref{p<p_e}), this yields $p=p_e$.

\section{Proof of $\langle \beta | \hat{M}_x |\beta \rangle =KL$ for $M=2^{mL}$}
\label{appendix_matsuzaki}
In this appendix, we prove 
$\langle \beta | \hat{M}_x |\beta \rangle \leq K L$ $(0<K<1)$ 
for $M=2^{mL}$ where $K$ is independent of $L$.
We use the following lemma:\\
{\it {Lemma}}:
There exists a real number $K$ independent of $L$ such that
 $2^{(1-m)} (1-K)^{1-K} K^{K}>1$ and $\frac{1}{2}<K<1$.\\
{\it {Proof}}:
Consider the following function:
\begin{eqnarray}
f(k) &=& (1-m)\log 2 + k\log k 
\nonumber\\
&&
+(1-k)\log (1-k)\ \ (\frac{1}{2}\leq k<1),
\end{eqnarray}
where $0<m<1$.  Since $f(\frac{1}{2})=-m \log 2 <0$,
 $f(k)\rightarrow (1-m)\log 2 >0\ (k\rightarrow 1)$, and the function $f(k)$
is continuous at $\frac{1}{2}\leq k<1$, 
we can apply the intermediate-value theorem,
therefore, there exists a real number $K$ such that
$f(K) > 0$ and $1/2<K<1$.
$\blacksquare$

Now, we prove $\langle \beta | \hat{M}_x |\beta \rangle =kL\
(0<k<1)$ for $M=2^{mL}$.
Let $P(M_x = -L+2j)$ be the probability of getting 
a value $-L+2j$ when one measures $\hat{M}_x$, where
$j=0,1,\cdots ,L$. It is represented as
\begin{eqnarray}
P(M_x = -L+2j) &=& 
\sum_{\nu }\big{|}\langle M_x=-L+2j,  \nu |\beta \rangle
  \big{|}^2
\nonumber\\
&\leq& {L \choose j}\max _{\nu }[\ \big{|}\langle M_x=-L+2j,  \nu
   |\beta \rangle \big{|}^2\ ],
\nonumber\\
\end{eqnarray}
where $|M_x=L-j, \nu \rangle $ is an eigenstate of $\hat{M}_x$, 
and $\nu =1,2,\cdots , {L \choose j}$ labels degenerate eigenstates.
Since
\begin{eqnarray}
\big{|}\langle M_x=-L+2j,  \nu
   |\beta \rangle \big{|} 
&\leq& 
\frac{1}{\sqrt{M}}\sum_{n=1}^{M}\big{|}\langle M_x=-L+2j,  \nu
  |x_n \rangle \big{|}
\nonumber\\
&\leq& \sqrt{\frac{M}{2^L}},
\end{eqnarray}
we have, 
using Stirling's formula $n! \sim \sqrt{2\pi n}(n/e)^n$,  
\begin{eqnarray}
&&
P(M_x = -L+2j)
\nonumber\\
&& \leq 
{L \choose j}\frac{M}{2^L}
\nonumber\\
&& \sim
\frac{1}{\sqrt{2\pi
  j(1-\frac{j}{L})}\{2^{(1-m)}(1-\frac{j}{L})^{1-\frac{j}{L}}
  (\frac{j}{L})^{\frac{j}{L}}
  \}^L}.\label{evaluation-P}
\end{eqnarray}
%
%
From the above Lemma, 
there exists a real number $K$ independent of
 $L$ such that
 $2^{(1-m)}(1-K)^{1-K}K^{K}>1$ and $\frac{1}{2}<K<1$.
Hence, 
\begin{eqnarray}
 L^nP(M_x = -L+2j) \rightarrow 0 \ \ (KL\leq j\leq L)
  \ \text{ as }L\rightarrow \infty ,\label{approach-to-zero}
\end{eqnarray}
 where $n$ is independent of $L$.
This yields 
\begin{equation}
\langle \beta |\hat{M}_x|\beta \rangle \to 
\sum_{j=0}^{KL-1}(-L+2j)P(M_x=-L+2j)
\end{equation}
%
%
as $L\rightarrow \infty $,
from which we conclude that 
$\langle \beta |\hat{M}_x|\beta \rangle \leq KL$ for large $L$.
$\blacksquare$

\section{Maximum eigenvalue of the VCM for product states}\label{app:emax.product.state}

In this appendix, we show that $e_{\rm max} = 2$ for
a product state, 
$ 
|\psi\rangle
=
\bigotimes_{l=1}^L |\phi_l \rangle_l,
$ 
where $|\phi_l \rangle_l$ denotes a state of 
the qubit at site $l$
($=1, 2, \cdots, L$).
The VCM of such a state is block-diagonal:
\begin{equation}
\left(
\begin{array}{ccccc}
\bm{V}_1 & 0 & \cdots & \cdots & 0\\
0 & \bm{V}_2 & 0 & \cdots & \cdots\\
\vdots & 0 & \bm{V}_3 & 0 &\cdots \\
\vdots & \vdots & & \cdots & \cdots \\
0 & \cdots & \cdots & 0 & \bm{V}_L 
\end{array}
\right),
\end{equation}
where $\bm{V}_l$ is a $3 \times 3$ matrix 
whose $\alpha \beta$ element ($\alpha, \beta = x, y, z$) is given by
\begin{equation}
(V_l)_{\alpha \beta}
=
\langle \phi_l|\hat{\sigma}_\alpha(l)\hat{\sigma}_\beta(l)|\phi_l\rangle
-\langle \phi_l|\hat{\sigma}_\alpha(l)|\phi_l\rangle\langle\phi_l|\hat{\sigma}_\beta(l)|\phi_l\rangle.
\end{equation}
Therefore, $e_{\rm max}$ is given by 
the maximum one among the eigenvalues of 
$\bm{V}_l$'s.
By a unitary transformation of this $3 \times 3$ matrix
such that $|\phi_l\rangle$ becomes an eigenstate 
of the transformed $\hat \sigma_z(l)$, 
we can transform $\bm{V}_l$ into 
\begin{equation}
\bm{V}_l=
\left(
\begin{array}{ccc}
1 & i & 0\\
-i & 1 & 0 \\
0 & 0 & 0   
\end{array}
\right).
\end{equation}
Since the maximum eigenvalue of this matrix is $2$, 
we find that $e_{\rm max} = 2$.


\begin{thebibliography}{99}
\bibitem{note:a(l)}
For homogeneous states, it is evident that $\hat A(L)$ should be 
homogeneously extended with increasing $L$.
For example, if $\hat A(L)=\sum_{l=1}^L \hat \sigma_z(l)$ for some $L$, 
then $\hat A(L')$ for another $L'$ ($\neq L$) should be 
$\sum_{l=1}^{L'} \hat \sigma_z(l)$.
For this reason, the condition $\| \hat a(l) \|=1$ was not imposed
in refs.~\citen{SM02,MSS05} or \citen{MS06}.
On the other hand, we do {\em not} require here that $\hat a(l)$ is 
independent of $L$, because the independence is not well-defined
when treating general families of states.
\bibitem{2p-corr}
One can also apply the theories in refs.~\citen{SM02} and \citen{SM05}
to $n$-point correlations,
if $n$ $(\geq 3)$ is independent of the system size $L$
and if the $n$ points can be divided into two groups 
$1$ and $2$
in such a way that 
the maximum distance $\ell_i$ ($i=1,2$) between two points in
each group is independent of $L$.
In fact, let $\ell \equiv \max_i \ell_i$, and 
regard each $\ell$ neighboring sites as a new site.
The original lattice composed of $L$ sites then becomes 
a new lattice composed of $L/\ell$ new sites.
Then, the original $n$-point correlations become two-point correlations
of the new lattice, and the theories of 
refs.~\citen{SM02} and \citen{SM05} become applicable
to the new lattice. 
By interpreting the results in terms of the original sites, 
one obtains the corresponding results for
the $n$-point correlations.
\bibitem{rough}
Strictly speaking, this picture is not necessarily correct because, 
when quantum states are strongly inhomogeneous,
$L e_{\rm max}$ is not necessarily 
proportional to the number of pairs of sites whose
correlation is of $\overline{\Theta}(1)$.
Fortunately, however, 
this picture is good 
for most instances in our simulations.
\end{thebibliography}
\end{document}